\documentclass[fleqn,usenatbib]{mnras}

\usepackage{ulem}
\usepackage{listings}
\usepackage{hyperref}
\hypersetup{
    colorlinks=true,
    linkcolor=blue,
    filecolor=magenta,      
    urlcolor=cyan,
}

\usepackage[T1]{fontenc}

\DeclareRobustCommand{\VAN}[3]{#2}
\let\VANthebibliography\thebibliography
\def\thebibliography{\DeclareRobustCommand{\VAN}[3]{##3}\VANthebibliography}

\usepackage{graphicx}	
\usepackage{amsmath}	
\usepackage{amssymb}	
\usepackage{newtxtext,newtxmath}

\title[Few-body toolkit SpaceHub]{SpaceHub:  A high-performance gravity integration toolkit for few-body problems in astrophysics}

\author[Wang et al.]{
Yi-Han Wang,$^{1}$\thanks{E-mail: yihan.wang.1@stonybrook.edu}
Nathan  W. C. Leigh,$^{3,4}$
Bin Liu,$^{5,6}$
Rosalba Perna$^{1,2}$
\\
$^{1}$Department of Physics and Astronomy, Stony Brook University, Stony Brook, NY, 11794, USA\\
$^{2}$Center for Computational Astrophysics, Flatiron Institute, 162 5th Avenue, New York, NY 10010, USA\\
$^{3}$Departamento de Astronomia, Facultad de Ciencias Fisicas y Matematicas, Universidad de Concepcion, Concepcion, Chile\\
$^{4}$Department of Astrophysics, American Museum of Natural History, Central Park West and 79th Street, New York, NY 10024\\
$^{5}$Niels Bohr International Academy, Niels Bohr Institute, Blegdamsvej 17, 2100 Copenhagen, Denmark\\
$^{6}$Cornell Center for Astrophysics and Planetary Science, Department of Astronomy, Cornell University, Ithaca, NY 14853, USA\\
}

\date{Accepted XXX. Received YYY; in original form ZZZ}

\pubyear{2015}

\begin{document}
\label{firstpage}
\pagerange{\pageref{firstpage}--\pageref{lastpage}}
\maketitle

\begin{abstract}
We present the open source few-body gravity integration toolkit {\tt SpaceHub}. {\tt SpaceHub} offers a variety of algorithmic methods, including the unique algorithms AR-Radau, AR-Sym6, AR-ABITS and AR-chain$^+$ which we show out-perform other methods in the literature and allow for fast, precise and accurate computations to deal with few-body problems ranging from interacting black holes to planetary dynamics. We show that AR-Sym6 and AR-chain$^+$,  with algorithmic regularization, chain algorithm, active round-off error compensation and a symplectic kernel implementation, are the fastest and most accurate algorithms to treat black hole dynamics with extreme mass ratios, extreme eccentricities and very close encounters. AR-Radau, the first regularized Radau integrator with round off error control down to 64 bits floating point machine precision, has the ability to handle extremely eccentric orbits and close approaches in long-term integrations. AR-ABITS, a bit efficient arbitrary precision method, achieves any precision with the least CPU cost compared to other open source arbitrary precision few-body codes. With the implementation of deep numerical and code optimization, these new algorithms in {\tt SpaceHub} prove superior to other popular high precision few-body codes in terms of performance, accuracy and speed. 
\end{abstract}




\section{Introduction}

Few-body gravity integrators are fundamental to the study of black hole and planetary dynamics.  The resulting simulations capture the time evolution of the orbital dynamics, evolving the system forward in time due to pair-wise gravitational interactions.  One of the primary challenges these integrators face is accurately capturing the integration of eccentric orbits and close approaches between particles.  This is because the integrations require extreme accuracy and precision in the vicinity of the singularity, where the distance $r$ between particles  becomes very small, which is made challenging due to the $1/r^2$ scaling characteristic of the Newtonian gravitational acceleration.  The most obvious solution is to choose a very small time step $r\rightarrow0$, but this can be very time consuming computationally. 

The introduction of regularization techniques to handle the computations in the vicinity of the singularity revolutionized the field of gravitational dynamics.  This stimulated the development of a number of mathematical transformations that can be implemented to improve accuracy and precision.  One prominent example includes Kustaanheimo-Stiefel (KS) regularization and applying the KS transformation to the perturbed two-body problem. \citet{Aarseth1974,Heggie1974,Zare1974} then introduced this transformation into general N-body problems, and \citet{Levison1994} applied it to planetary dynamics. Later on, new Logarithmic Hamiltonian regularization with a leapfrog scheme was invented by \citet{Mikkola1999a,Mikkola1999b} and \citet{Preto1999}. This method can accurately and efficiently capture the dynamics in the vicinity of the singularity using a regularized equation of motion, but the method becomes invalid in the limit of extremely large mass ratios. To overcome this challenge, \citet{Mikkola2002} introduced the time-transformed leapfrog scheme, which allows for an arbitrary regularization mass coefficient function to deal with extreme mass ratios during close approaches. 

All regularized methods that include the algorithmic regularization introduced by \citet{mikkola_implementing_2008} are based on the leapfrog scheme, since it can maintain the symplectic nature of the system when adaptive stepping is not implemented. Because the method is symplectic, it adheres to Hamilton's equations and time-reversibility is preserved.  However, the leapfrog method is only a second order method, and going to higher orders can greatly improve accuracy and precision. Consequently, the rational Bulirsch-Stoer(BS) extrapolation \citep{Gragg1965,Press1986} was introduced into the regularization method. This extrapolation method can efficiently construct higher order methods that adhere to the leapfrog scheme. With the BS extrapolation and regularization techniques now developed, efficient high precision integrations can be performed that accurately treat extremely eccentric orbits and very close pair-wise particle approaches.  These problems correspond to prompt events, which may or may not occur repeatedly.  But what if long term integrations are required, where such close pair-wise approaches \textit{do} occur regularly?  

Indeed, the computations become more challenging as longer time scale integrations are needed.  Examples include extremely eccentric systems that require integrating over many orbits, such as the gravitational wave-induced inspiral of eccentric binaries down to the kilo-Hertz level, high eccentricity tidal dissipation, and so on.  The reason problems arise is because, for integration methods based on extrapolation, the higher order extrapolation reduces the truncation error but the round off error becomes significant due to the finite bit floating-point number arithmetic. The resulting large round off errors can severely reduce the precision and destroy the long term performance of high precision methods. Therefore, reducing and properly managing round off errors is essential in high precision integration methods. 
 
 For systems where extreme approaches between particles arise, the close value subtraction of the floating-point numbers that define the positions of the particles causes several significant digits to be lost.  This introduces round off errors into the integrations. To solve this problem, a chain coordinate system \citep{Mikkola1993} was invented to evaluate the relative distances during close encounters. With this, the relative distances between particles are directly replaced by the chain coordinates, thus avoiding the close value subtraction. However, the chain coordinates need to be reconstructed and updated after each time step. The problem that arises here is that frequent chain coordinate updates involve large quantities of floating-point arithmetic and introduces extra round off errors into long term integrations. Branched tree coordinates are more optimized for maintaining the shortest relative position between particles and can alleviate this problem. \citet{rantala_mstar_2020} implemented the minimum spanning tree in their AR-Chain based code MSTAR, and showed that branched tree coordinates are more efficient, with less round off error accumulation than chain coordinates in the large-N ($\sim$ 400) regime.
 
 Close value subtractions are not the only source of round off errors that can affect long term integration performance. In BS-based methods, round off errors propagate and can be magnified through the extrapolation table.  With higher order extrapolations, the round off errors become more significant. Theoretically, the BS extrapolation can achieve arbitrarily high order, but it can only achieve 14-16th order for double precision floating-point numbers. Consequently, even if the BS extrapolation is very efficient in achieving higher order symplectic methods such as the leapfrog scheme, the precision is limited by the extrapolation. More importantly, the extrapolation process is not time symmetric, which breaks the symplectic nature of the system even if fixed time steps are adopted. Therefore, alternative methods are needed to achieve higher order results and maintain time symmetry.

For long term integrations, where many time steps are needed, another source of round off error  becomes non-negligible. During the integration, small incremental steps can be used to better evolve specifically physical quantities in each step. But this can introduce large difference additions between floating-point numbers, which will in turn introduce significant round off errors into the integration due to the finite bit truncation. This kind of round off error accumulates continuously over the course of the integration. To alleviate this problem, \citet{quinn_roundoff_1990} proposed an active round off error compensation method in floating-point number arithmetic. This was introduced into the Gauss-Radau \citep{everhart_efficient_1985} scheme by \citet{rein_ias15_2014}, which proved useful in slowing down the round off error accumulation in long term integrations. However, the raw Gauss-Radau method is cumbersome in dealing with extreme eccentricities and very close pair-wise encounters, causing the integrations to lose significant precision for these specific problems.

In this paper, we present a new few-body gravity integration toolkit called {\tt SpaceHub}. It contains various previously developed algorithms in addition to several new and novel techniques which improve the speed, accuracy and precision of the calculations even further.  This includes state-of-the-art algorithms that can efficiently deal with both extreme eccentricities and extreme mass ratios even for long term integrations. Problems which have already been investigated using this code include interactions between compact object binaries and supermassive black hole binaries \citep{Wang2018,Wang2019a,Wang2019b,Liu2019a,Liu2019b}, formation of black hole binaries in dense star clusters \citep{Perna2019}, interactions between stars and planetary systems
\citep{Wang2020a,Wang2020b,Wang2020c}.

The paper is organized as follows.  In Section 2, we introduce and briefly review several previously developed few body integration techniques, including the regularization algorithm, the chain algorithm, active error compensation and the Bulirsch-Stoer extrapolation. We explicitly highlight our improvements to those techniques, and introduce three new methods in {\tt SpaceHub} to accomplish this goal. In Section 3, we discuss how {\tt SpaceHub} can easily construct new algorithms and our optimization implementations to improve performance. In section 4, we present various tests of our newly developed algorithms in {\tt SpaceHub}, and compare to other popular high precision few body codes. We summarize our main results and present our conclusions in Section 5.

\section{Review of High precision few-body integration methods}
There are several integration methods, including AR-Chain, IAS15 and {\tt Brutus}, that can achieve very high precision with low relative energy errors. In this section we will review these integration schemes and point out any shortcomings of the integration methods. This will help provide a broader context  to then present the improvements  made by {\tt SpaceHub} which will be discussed in Sec.~3.  

\subsection{AR-chain}

For the N-body problem in the small-N regime, stochastic close pair-wise approaches and high eccentricity orbits in collisional systems can significantly hurt the integration accuracy. Due to the strong interaction around the closest approach, very small time steps are required to properly resolve the dynamics in the vicinity of the singularity in conventional step control methods. But increasing the number of time steps severely slows down the integration and, especially for longer term integrations, more floating point arithmetic introduces more round off errors. In the past few decades, several methods have been invented to tackle this problem, including KS regularization, the Logarithmic Hamiltonian method (LogH) \citep{Mikkola1999a,Mikkola1999b, Preto1999}, the time-transformed leapfrog (TTL) regularization scheme\citep{Mikkola2002} and generalized midpoint regularization (GAR)\citep{stetter_1968}. By transforming the equations of motion, the combination of all of these regularization methods can efficiently resolve close pair-wise approaches and high eccentricity orbits. \citet{mikkola_implementing_2008} gathered all the three regularization schemes (LogH, TTL and GAR) to form the Algorithmic Regularization scheme in his AR-chain algorithm to deal with velocity-dependent problems for extreme eccentricity cases and close pair-wise approaches between particles. Here we briefly review the regularization methods implemented in AR-chain.

\subsubsection{LogH regularization}
For systems with equations of motion for each particle,
\begin{eqnarray}
    \frac{d\mathbf{x}_i}{dt} &=& \mathbf{v}_i\\
    \frac{d\mathbf{v}_i}{dt} &=& \mathbf{g}_i + \mathbf{f}_{i}
\end{eqnarray}
where $\mathbf{x}_i$, $\mathbf{v}_i$, $\mathbf{g}_i$ and $\mathbf{f}_i$ are, respectively, the position, velocity, Newtonian acceleration and other accelerations in addition to the Newtonian one acting on particle $i$. The LogH regularization scheme introduces additional quantities, namely $T$, $U$ and $B$, to transform the equations of motion during the integration. Here, $T$ is the total kinetic energy of the system, $U$ is the absolute value of the total Newtonian potential energy of the system and $B$ is the binding energy of the system,
\begin{eqnarray}
    T &=& \sum_i \frac{1}{2}m_i \mathbf{v}_i^2\,,\\
    U &=& \sum_{i<j} \frac{m_i m_j}{|\mathbf{x}_i - \mathbf{x}_j|}\,,\\
    B &=& U-T\,.\label{eq:b}
\end{eqnarray}
Thus, the regularized algorithm can be described in the time-symmetric leap-frog form as follows

$\mathbf{D}(h)$:
\begin{eqnarray}
 T &=& \sum_i \frac{1}{2}m_i \mathbf{v}_i^2\\
 dt &=&  h/(T+B)\\
 t  &\rightarrow& t + dt\\
 \mathbf{x}_i &\rightarrow& \mathbf{x}_i + \mathbf{v}_i dt
\end{eqnarray}

$\mathbf{K}(h)$:
\begin{eqnarray}
 U &=& \sum_{i<j} \frac{m_i m_j}{|\mathbf{x}_i - \mathbf{x}_j|}\\
 dt &=&  h/U\\
 B &\rightarrow& B - \frac{dt}{2}\sum_i m_i \mathbf{v}_i\cdot \mathbf{f}_{i} \\
 \mathbf{v}_i &\rightarrow& \mathbf{v}_i + (\mathbf{g}_i + \mathbf{f}_{i}) dt\\
 B &\rightarrow& B - \frac{dt}{2}\sum_i m_i \mathbf{v}_i\cdot\mathbf{f}_{i} 
\end{eqnarray}
Then, a complete one step integration can be constructed as $\mathbf{D}(h/2)\mathbf{K}(h)\mathbf{D}(h/2)$ or $\mathbf{K}(h/2)\mathbf{D}(h)\mathbf{K}(h/2)$, where $h$ is the time step.

\subsubsection{TTL regularization}
For systems with extreme particle mass ratios, the regularization function $U=\sum_{i<j}\frac{m_im_j}{|\mathbf{x}_i-\mathbf{x}_j|}$ and $T+B$ in the LogH method can end up being dominated by the most massive pair of particles in the system. Therefore, even if close encounters between extreme mass ratio pairs occur during the integration, it can happen that the step size is not being properly regularized.

In order to compensate for this, instead of using $U$ and $T+B$ as the regularization functions, \citet{Mikkola2002} proposed two independent variables $\omega$ and $\Omega$ to perform the require transformations in the equations of motion,
\begin{eqnarray}
 \Omega &=& \sum_{i<j}\frac{\Omega_{ij}}{|\mathbf{x}_i - \mathbf{x}_j|}\\
\frac{d\omega}{dt} &=& \sum_i \frac{\partial \Omega}{\partial \mathbf{x}_i}\cdot \mathbf{v}_i, \omega(0) = \Omega(0)\,,
\end{eqnarray}
where $\Omega_{ij}$ is a function of the particle masses. Conventionally, $\Omega_{ij}=m_im_j$ can be adopted, which is mathematically equivalent to the LogH method.  However, alternative choices can be made, such as the mass averaged function,
\begin{eqnarray}
 \Omega_{ij}&=&
 \left\{
 \begin{array}{ll}
      \tilde{m}^2  \quad\text{if $m_im_j<\epsilon \tilde{m}^2$}\\
      0  \quad\quad \text{otherwise}\,.
    \end{array}   
\right.
\end{eqnarray}
 With this kind of regularization function, the contribution from small mass pairs becomes more dominant in the regularization function such that close encounters between low-mass particles can be correctly captured. Here,  $\tilde{m}^2 = \sum_{i<j}2m_im_j/(N(N-1))$ is the mean mass of the system with N particles and $\epsilon$ is a parameter to set the threshold where only particle pairs with mass products small enough can contribute to the regularization function. In this way, large mass pairs do not contribute to the regularization; hence the contribution of the small mass pairs in the regularization function will not be overwhelmed due to the contribution from large mass pairs. \citet{Mikkola2002} suggested the value of $\epsilon$ to be  $10^{-3}$ in order to correctly capture the small mass pair interactions. 
 
 The leap-frog scheme with TTL regularization can be written in form of 
$\mathbf{D}(h)$:
\begin{eqnarray}
dt &=&  h/\omega\\
 t  &\rightarrow& t + dt\\
 \mathbf{x}_i &\rightarrow& \mathbf{x}_i + \mathbf{v}_i dt
\end{eqnarray}

$\mathbf{K}(h)$:
\begin{eqnarray}
 \Omega &=& \sum_{i<j}\frac{\Omega_{ij}}{|\mathbf{x}_i - \mathbf{x}_j|}\\
 dt &=&  h/\Omega\\
 \omega &\rightarrow& \omega + \frac{dt}{2}\sum_i \frac{\partial \Omega}{\partial \mathbf{x}_i}\cdot \mathbf{v}_i\\
 \mathbf{v}_i &\rightarrow& \mathbf{v}_i + (\mathbf{g}_i + \mathbf{f}_{i} ) dt\\
 \omega &\rightarrow& \omega + \frac{dt}{2}\sum_i \frac{\partial \Omega}{\partial \mathbf{x}_i}\cdot \mathbf{v}_i
\end{eqnarray}
with $\omega(0)=\Omega(0)$.

\subsubsection{Generalized midpoint method}
To construct a leapfrog scheme, the variables on the right side of the equations of motion should be independent of the variables on the left side. However, if the external acceleration $\mathbf{f}_{i}$ is velocity dependent, then the kick step becomes problematic.  In this case, $\mathbf{v}_i$ needs implicit iteration on both sides of the equation  which breaks the time symmetry of the leapfrog scheme,
\begin{equation}
    \mathbf{v}_i \rightarrow \mathbf{v}_i + (\mathbf{g}_i + \mathbf{f}_{i}(\mathbf{v}_i,...) ) dt\,.\\
\end{equation}
In order to preserve the time symmetry of the leapfrog scheme, \citet{stetter_1968, Mikkola2002} introduced the pseudo-velocity $\mathbf{w}_i$ in the kick-step procedure in order to to maintain the time symmetry. For systems with equations of motion of the form
\begin{eqnarray}
    \frac{d\mathbf{x}_i}{dt} &=& \mathbf{v}_i\\
    \frac{d\mathbf{v}_i}{dt} &=& \mathbf{g}_i + \mathbf{f}_{i}(\mathbf{x}) + \mathbf{f}_{i,v}(\mathbf{x},\mathbf{v})
\end{eqnarray}
where $ \mathbf{f}_{i}(\mathbf{x})$ is the external velocity-independent acceleration and $\mathbf{f}_{i,v}(\mathbf{x},\mathbf{v})$ are the external velocity-dependent accelerations, the leapfrog scheme can be constructed in the following way for the LogH method,

$\mathbf{D}(h)$:
\begin{eqnarray}
 T &=& \sum_i \frac{1}{2}m_i \mathbf{v}_i^2\\
 dt &=&  h/(T+B)\\
 t  &\rightarrow& t + dt\\
 \mathbf{x}_i &\rightarrow& \mathbf{x}_i + \mathbf{v}_i dt
\end{eqnarray}

$\mathbf{K}(h)$:
\begin{eqnarray}
 U &=& \sum_{i<j} \frac{m_i m_j}{|\mathbf{x}_i - \mathbf{x}_j|}\\
 dt &=&  h/U
\end{eqnarray} 
 
 \indent\indent $\mathbf{K}_v(dt/2)$:
 \begin{eqnarray}
  \mathbf{v}_i &\rightarrow& \mathbf{v}_i + (\mathbf{g}_i + \mathbf{f}_{i}+\mathbf{f}_{i,v}(\mathbf{x},\mathbf{w})) \frac{dt}{2}
 \end{eqnarray}
  
  \indent\indent $\mathbf{K}_w(dt)$:
  \begin{eqnarray}
   \mathbf{w}_i &\rightarrow& \mathbf{w}_i + (\mathbf{g}_i + \mathbf{f}_{i}+\mathbf{f}_{i,v}(\mathbf{x},\mathbf{v}))dt\\
   B &\rightarrow& B - dt\sum_i m_i \mathbf{v}_i\cdot(\mathbf{f}_{i}+ \mathbf{f}_{i,v}(\mathbf{x},\mathbf{v})) 
  \end{eqnarray}
  
   \indent\indent $\mathbf{K}_v(dt/2)$:
\begin{eqnarray}
  \mathbf{v}_i &\rightarrow& \mathbf{v}_i + (\mathbf{g}_i + \mathbf{f}_{i}+\mathbf{f}_{i,v}(\mathbf{x},\mathbf{w})) \frac{dt}{2}
 \end{eqnarray}
with $\mathbf{w}_i(0)$=$\mathbf{v}_i(0)$. Note that the sub-leapfrog step $\mathbf{K_v}(dt/2)\mathbf{K_w}(dt)\mathbf{K_v}(dt/2)$ can also be constructed as $\mathbf{K_w}(dt/2)\mathbf{K_v}(dt)\mathbf{K_w}(dt/2)$. The same scheme can also be obtained for the TTL method with external velocity-dependent accelerations.

\subsubsection{Chain coordinates}\label{sec:chain-coord}
The AR-chain method was invented to deal with close encounters and extremely eccentricity orbits, thus the relative positions between particles in the integrated system can be very small. To evaluate the acceleration between close particles, the relative position between two close particles is needed. This calculation requires a close value subtraction arithmetic between two floating-point numbers, which can cause fast round off error accumulation. To solve this problem \citet{mikkola_chain_1990, Mikkola1993} introduced the chain coordinates into the few-body integration, where the relative positions between particles are calculated from the initial conditions and converted to chain coordinates.  Instead of evolving the original Cartesian coordinates, the relative positions (i.e., chain coordinates) will be evolved.  
{Since the chain coordinates are constructed in a way that the shortest few relative positions are always kept in the chain, the close value subtraction between close positions can be replaced by the chain coordinates directly. This coordinate transformation can reduce the round off errors from close value subtractions between floating-point numbers, thus giving better error control in the integration. 
The transformation from Cartesian coordinates to chain coordinates can be written as 

\begin{eqnarray}\label{eq:chain}
 \mathbf{X}_k = \mathbf{x}_{i_{k+1}}-\mathbf{x}_{i_k}\\
 \mathbf{V}_k = \mathbf{v}_{i_{k+1}}-\mathbf{v}_{i_k}
\end{eqnarray}
with $k = 1,2,...,N-1$ and $i$ is the index of the Cartesian coordinates in chain coordinates. Then the inverse transformation can be written as
\begin{eqnarray}
 \tilde{\mathbf{x}}_{i_1} &=& \mathbf{0}\\
 \tilde{\mathbf{x}}_{i_{k+1}} &=& \tilde{\mathbf{x}}_{i_{k}} + \mathbf{X}_{k}\\
 \tilde{\mathbf{v}}_{i_1} &=& \mathbf{0}\\
 \tilde{\mathbf{v}}_{i_{k+1}} &=& \tilde{\mathbf{v}}_{i_{k}} + \mathbf{V}_{k}
\end{eqnarray}
followed by a reduction to the centre of mass reference frame
\begin{eqnarray}
 \tilde{\mathbf{x}}_{\rm cm} &=& \sum_j m_j \tilde{\mathbf{x}}_j/\sum_j m_j\\
 \tilde{\mathbf{v}}_{\rm cm} &=& \sum_j m_j \tilde{\mathbf{v}}_j/\sum_j m_j\\
 \mathbf{x}_j &=& \tilde{\mathbf{x}}_j - \tilde{\mathbf{x}}_{\rm cm}\\
 \mathbf{v}_j &=& \tilde{\mathbf{v}}_j - \tilde{\mathbf{v}}_{\rm cm}.
\end{eqnarray}
The equations of motion of the chain coordinates are then
\begin{eqnarray}\label{eq:chain}
  \frac{d\mathbf{X}_k}{dt} &=& \mathbf{V}_k,\quad\quad k=1,2,...,N-1\\
    \frac{d\mathbf{V}_k}{dt} &=& \mathbf{g}_{i_{k+1}}- \mathbf{g}_{i_k} + \mathbf{f}_{i_{k+1}}-\mathbf{f}_{i_{k}} + \mathbf{f}_{i_{k+1},v}-\mathbf{f}_{{i_k},v}\,.
\end{eqnarray}
With the new equations of motion, every position subtraction $\mathbf{r}_{jk}$ between $\mathbf{x}_j$ and $\mathbf{x}_k$ (where $j$ and $k$ are the chain coordinates indices) is
\begin{eqnarray}
 \mathbf{r}_{jk} = 
 \left\{
 \begin{array}{lr}
      \pm(\mathbf{x}_k - \mathbf{x}_j)\quad&  \text{if $k>j\pm2$ (far pair)} \\
       \pm\mathbf{X}_j \quad &\text{if $k=j\pm1$ (close pair)}\\
       \pm(\mathbf{X}_j + \mathbf{X}_{j+1})\quad &\text{if $k=j\pm2$ (intermediate pair)}\,.
 \end{array}
 \right.
\end{eqnarray}
}
\subsubsection{Gragg-Bulirsch-Stoer extrapolation}
The algorithmic regularization and chain coordinates can be constructed in the form of the leapfrog scheme. However, by itself, the leapfrog method is only a two order method.  This makes it inefficient in achieving high precision. To tackle this problem and achieve high precision, higher order methods are usually needed. The Gragg-Bulirsch-Stoer extrapolation \citep{Gragg1965,Bulirsch1966,Press1986} can construct higher order results from a series of lower order results through an extrapolation table:
\begin{table}
\begin{center}
\begin{tabular}{ |l|l|l|l|l|l|l|l| } 
$T_{11}$ &&&&&&&\\
&$\nwarrow$&&&&&&\\
$T_{21}$ & $\leftarrow$ & $T_{22}$ & &&&&\\
& $\nwarrow$ &&$\nwarrow$ &&&&\\
$T_{31}$ & $\leftarrow$ & $T_{32}$ &  $\leftarrow$ & $T_{33}$&&&\\
& $\nwarrow$ &&$\nwarrow$ &&$\nwarrow$&&\\
$T_{41}$ & $\leftarrow$ & $T_{42}$ & $\leftarrow$ & $T_{43}$& $\leftarrow$&$T_{44}$\\
...& & ...&& ...&& ...&
\end{tabular}
\end{center}
\caption{Gragg-Bulirsch-Stoer extrapolation table.\label{tab:BS}}
\end{table}
For integrations with macro step $H$, the first column of Table~\ref{tab:BS} will be filled with basic lower order methods with $n_i$ sub-steps $h_{n_i}=H/n_i$. Then higher order results can be constructed recursively by extrapolation,
\begin{equation}\label{eq:BSextrap}
    T_{i,j} = T_{i, j-1} + \frac{T_{i,j-1} - T_{i-1, j-1}}{(n_i/n_{i-j+1})^2-1}.
\end{equation}
If the basic lower order method (such as the leapfrog scheme) is only in even powers of the time-step, the final extrapolated result $T_{k,k}$ has $p+2(k-1)$-th order precision, where $p$ is the order of the basic integration method used in the first column. If the basic integration method is non-symmetric in time, then $T_{k,k}$ has $p+k-1$-th order precision. Therefore, GBS extrapolation is efficient in constructing higher order methods for time symmetric integration schemes like leapfrog. Since the regularization chain algorithm can be written in the form of a time-symmetric leapfrog scheme, where all odd powers of the  time-steps vanish, the GBS extrapolation is adopted in the original AR-chain \citep{mikkola_implementing_2008} to achieve higher order accuracy and precision.

\subsection{IAS15}
IAS15 ({\tt Rebound}, \citealt{rein_ias15_2014}) is a Gauss-Radau based integration method with improvements on step size control and round off error reduction. For few-body problems without high eccentricities and close pair-wise encounters, it performs very well in terms of maintaining low relative energy errors and minimizing long term unbiased round off error accumulation \citep{Brouwer1937,Brouwer2013}. In this subsection, we will briefly introduce the the Gauss-Radau method and the main improvements implemented in IAS15. 

\subsubsection{Gauss-Radau integration}
\citet{everhart_efficient_1985} introduced a modified 15th order Runge-Kutta method with Gauss-Radau spacings to achieve very high precision in few-body problems. This method solves the general equation
\begin{equation}
    y^{\prime\prime} = F(y, y^\prime, t).
\end{equation}
In celestial mechanics, $y$ denotes the positions of the particles, thus $F$ denotes the accelerations. \citet{everhart_efficient_1985} expanded the equation into a truncated Taylor series,
\begin{equation}
     y^{\prime\prime}|_{h} \sim y^{\prime\prime}_0 + a_1t+a_2t^2+...+a_7t^7\,.
\end{equation}
Using the dimensionless time-step $h = t/dt$ and $b_i=a_it^i$, the equation can be rewritten as
\begin{equation}
     F = y^{\prime\prime}|_{h} = \sim y^{\prime\prime}_0 + b_1h+b_2h^2+...+b_7h^7\,.
\end{equation}
After $y^{\prime\prime}$ is found at $h_1$=0, the value of $b_i$ can be found recursively at each time step.  This is done by evaluating  $y^{\prime\prime}$ at a series of suitably chosen Gauss-Radau spacings $h_1, h_2,...h_8$ within the interval of $h$ between 0 and 1. Then, at the end of the macro step h, the accelerations $y^{\prime\prime}$ can be approximated by this Taylor series. The corresponding $y^\prime$ (velocity) and $y$ (position) values can thus be calculated analytically by integrating $y^{\prime\prime}$ over $dt$,
\begin{eqnarray}
    &&y^\prime|_{h} \sim y^\prime_0 + hdt\bigg(y^{\prime\prime}_0 +\frac{h}{2}\bigg(b_0+\frac{2h}{3}\bigg(b_1+...\bigg)\bigg)\bigg)\\
    &&y|_{h} \sim y_0 + y^\prime_0 hdt +\frac{h^2dt^2}{2}\bigg(y^{\prime\prime}_0 +\frac{h}{3}\bigg(b_0+\frac{h}{2}\bigg(b_1+...\bigg)\bigg)\bigg).
\end{eqnarray}
The $b_i$ are evaluated at each step by repeatedly iterating over the Gauss-Radau spacing $h_i$, where 
\begin{eqnarray}\label{eq:radau-b}
    b_1 &=& C_{71}g_7 + C_{61}g_6 + C_{51}g_5 + ...+ C_{11}g_1\nonumber\\
    b_2 &=&             C_{72}g_7 + C_{62}g_6 + ...+ C_{22}g_2\nonumber\\
    b_3 &=&                           C_{73}g_7+ ...+ C_{33}g_3\nonumber\\
    ...\nonumber\\
    b_7 &=&                                       C_{77}g_7
\end{eqnarray}
and 
\begin{eqnarray}\label{eq:radau-g}
    g_1 &=& (y^{\prime\prime}|_{h_2} - y^{\prime\prime}|_{h_1})R_{21}\\
    g_2 &=&  ((y^{\prime\prime}|_{h_3} - y^{\prime\prime}|_{h_1})R_{31} - g_1)R_{32}\nonumber\\
    g_3 &=&  (( (y^{\prime\prime}|_{h_4} - y^{\prime\prime}|_{h_1})R_{41} - g_1)R_{42} - g_2)R_{43}\nonumber \\
    ...\nonumber\\
    g_7 &=& ( ...(( (y^{\prime\prime}|_{h_8} - y^{\prime\prime}|_{h_1})R_{81} - g_1)R_{82} - g_2)R_{43}...-g_6)R_{87} \nonumber
\end{eqnarray}
where $C_{ij}$ and $R_{ij}$ are constants that can be calculated from $h_i$. This expansion of $y^{\prime\prime}$ up to $t^7$  with Gauss-Radau spacing has high precision, reaching 15th order.

\subsubsection{Active round off error reduction}
For higher order integration methods, the truncation error can be efficiently reduced by shrinking the step size. However, the round off error from the arithmetic of finite bit floating point numbers becomes dominant in high precision integration methods. Therefore, by increasing the order of the integration method or shrinking the step size we cannot obtain more precise solutions. On the contrary, shrinking the step size means more floating point arithmetic that will accumulate more round off errors, leading to less accurate solutions. For finite bit floating-point numbers with N mantissa bits, the relative number precision is 
\begin{equation}
    {\rm eps} = 2^{-(N-1)}\label{eq:eps}
\end{equation}
for double precision floating-point numbers with 53 mantissa bits, this is $\sim 2.2\times 10^{-16}$. Thus, for double precision floating-point number arithmetic, any relative value in the results smaller than ${\rm eps}$ will be rounded away.

 Other than the round off error from close value subtractions discussed in Section~\ref{sec:chain-coord}, another arithmetic operations can introduce large round off errors due to large difference additions, i.e., adding a small number to a large number. In large difference addition, the result of the calculation is of the same order as the large number. Hence, the rounding away aspect operates at the order of $\rm eps \times$ |large number|. Since the absolute value of the small number is much smaller than the large number, the rounded away part can be a significant component of the small number.  Consequently, the result loses considerable accuracy. 

Indeed, in numerical integrations, these large difference additions are needed in each step as we keeping evolving the variable $\xi$
\begin{equation}
    \xi \rightarrow \xi+ \frac{d\xi}{dt} dt
\end{equation}
where $\frac{d\xi}{dt} dt$ is usually small compared to $\xi$. 

To achieve higher precision, reducing the round off error becomes essential. \citet{quinn_roundoff_1990} introduced into celestial mechanics the active round off error reduction invented by \citet{Kahan1965}.  \citet{rein_ias15_2014} then implemented it in IAS15.

For floating-point number arithmetic, the result of each operation will be rounded adopting a certain round off strategy.  For example, consider the result of
\begin{equation}
    \xi+d\xi \rightarrow {\rm rnd}[\xi+d\xi]
\end{equation}
where $\rm rnd$ is a certain round off strategy. What is interesting is that the round off error cause by $\rm rnd$ can also be estimated under the floating-point number arithmetic, where
\begin{equation}
    {\rm err}_{+}(\xi, d\xi) \sim  {\rm rnd}[{\rm rnd}[{\rm rnd}[\xi+d\xi] - \xi ] - d\xi]
\end{equation}
Although,
\begin{equation}
    {\rm err}_{+}(\xi, d\xi) < {\rm eps} \times {\rm rnd}[\xi + d\xi]\,
\end{equation}
it can be comparable to $d\xi$. Since in the integration $d\xi$ at different $t$ will be continuously added to the $\xi$, the round off error from the last step can be compensated in the next step by subtracting the error from $d\xi$. A complete active round off error compensation process can be described as
\begin{eqnarray}\label{eq:active}
    \xi_{1} &=& {\rm rnd}[\xi_{0} + d\xi_{0}]\\
    {\rm err}_{+}(\xi_{0}, d\xi_{0}) &=&  {\rm rnd}[{\rm rnd}[\xi_1 - \xi_0 ] - d\xi_0]\\
    d\xi_1 &=& {\rm rnd}[ d\xi_1 - {\rm err}_{+}(\xi_{0}, d\xi_{0})]\\
    \xi_{2} &=& {\rm rnd}[\xi_{1} + d\xi_{1}]\\
    {\rm err}_{+}(\xi_{1}, d\xi_{1}) &=&  {\rm rnd}[{\rm rnd}[\xi_2 - \xi_1 ] - d\xi_1]\\
     d\xi_2 &=& {\rm rnd}[ d\xi_2 - {\rm err}_{+}(\xi_{1}, d\xi_{1})]\\
    ...\nonumber
\end{eqnarray}
By this means, the round off error can be reduced by 1-2 orders of magnitude.  This is essential in high precision integration methods trying to minimize round off errors.

\subsection{Arbitrary precision integration}
For integration methods with high precision, other than reducing the round off error with limited mantissa bit floating point numbers, another way to increase the precision of the integration method is to lower the $\rm eps$ in Equation~\ref{eq:eps}. The most straightforward way to do this is to increase the mantissa bit number N, i.e. use longer bit floating-point numbers.  In this way, one can achieve any precision by using longer and longer bit floating-point numbers. However, the length of the CPU register is limited.  Thus, the arithmetic between non-standard floating point numbers can be extremely slow compared to the arithmetic between standard floating point numbers.

Since the GBS extrapolation can simply construct arbitrarily higher orders, using longer bit floating point numbers in this method can easily achieve arbitrary precision. {\tt Brutus} \citep{Boekholt2015} implemented this method by adopting the arbitrary bit floating point number.

\section{{\tt SpaceHub} library}

In this section, we describe several novel features implemented in {\tt SpaceHub}, and quantify how each of these novel algorithms improve performance, in particular accuracy, precision and speed, relative to previous methods discussed in the previous section.  This is done by considering the time evolution of two example cases, namely the earth-moon-sun system and a highly eccentric binary composed of two solar mass stars with an initial  semi-major axis of 1 AU.

\subsection{AR-ABITS: Regularized arbitrary precision algorithm}\label{sec:arabits}
As with {\tt Brutus}, using extended mantissa bit floating-point numbers can reduce the $\rm eps$ in Equation~\ref{eq:eps}.  Thus, it can achieve arbitrary precision by increasing the bit length. However, the round off error will also propagate and accumulate through the extrapolation table in Table~\ref{tab:BS}. As the order of the GBS extrapolation increases, the round off will accumulate faster and faster. Thus the additional mantissa bits becomes less and less efficient in increasing the integration accuracy. Therefore, reducing the round off error in GBS extrapolation is also important in order to achieve the same precision with less mantissa bits and less CPU time. 

\subsubsection{Improvement on GBS extrapolation}

In the BS extrapolation, higher order methods can be obtained from lower order methods by extrapolation, as described in Equation~\ref{eq:BSextrap}. However, the round off error will also propagate and accumulate through the extrapolation table.
\begin{table}
\begin{center}
\begin{tabular}{ |r|r|r|r|r|r|r|r|r| } 
$\epsilon$ &&&&&&&& \\
-$\epsilon$ & -1.6$\epsilon$&&&&&&&\\
$\epsilon$ & 2.6$\epsilon$ & 3.1$\epsilon$&&&&&& \\
-$\epsilon$ & -3.6$\epsilon$ & -5.6$\epsilon$ & -6.2$\epsilon$ &&&&&\\
$\epsilon$ & 4.6$\epsilon$ & 9.1$\epsilon$ & 11.9$\epsilon$ & 12.7$\epsilon$ &&&&\\
-$\epsilon$ & -5.5$\epsilon$ & -13.6$\epsilon$ & -21.2$\epsilon$ & -25.3$\epsilon$ & -26.4$\epsilon$ &&&\\
$\epsilon$ & 6.5$\epsilon$ & 19.1$\epsilon$ & 35.0$\epsilon$ & 47.7$\epsilon$ & 54.1$\epsilon$ & 55.8$\epsilon$ &&\\
-$\epsilon$ & -7.5$\epsilon$ & -25.6$\epsilon$ & -54.3$\epsilon$ & -84.1$\epsilon$ & -105.6$\epsilon$ & -116.3$\epsilon$ & -119.0$\epsilon$ &\\
...& ...& ...&...& ...& ...& ...& ...&
\end{tabular}
\end{center}
\caption{Round off error propagation in the Bulirsch-Stoer extrapolation with step sequence $n_i = 2i$.\label{tab:roundoff1}}
\end{table}
If the round off error from the base integration method is $\epsilon, -\epsilon, \epsilon,...$ in the first column, then the round off error propagation with the extensively used step sequence $n_i = 2 i$ suggested by \citet{Deuflhard1983} will behave as in Table~\ref{tab:roundoff1}. As shown in the table, $T_{8,8}$ with 16th order precision magnified the round off error to become 100 times larger. Thus, we will lose at least 2 significant digits from the extrapolation. This becomes more problematic for higher order extrapolations, as required by arbitrary precision integration methods with extremely low tolerance.

To reduce the round off error in the extrapolation process, there are two fine-tunings that can be performed on the extrapolations. The first is a step sequence choice. The step sequences $n_i=1, 2, 3, 5, 8, 12, 17, 25, 36, 51, 73,...$\citep{Fukushima1996} is better than the extensively used sequence $n_i=2i$, where the extrapolation coefficient reduces the round-off error propagation. 
\begin{table}
\begin{center}
\begin{tabular}{ |r|r|r|r|r|r|r|r|r| } 
$\epsilon$ &&&&&&&&\\
-$\epsilon$ & -1.7$\epsilon$&&&&&&&\\
$\epsilon$ & 2.6$\epsilon$ & 3.1$\epsilon$&&&&&&\\
-$\epsilon$ & -2.1$\epsilon$ & -3.0$\epsilon$ & -3.3$\epsilon$&&&&&\\
$\epsilon$ & 2.3$\epsilon$ & 3.0$\epsilon$ & 3.4$\epsilon$ & 3.5$\epsilon$&&&&\\
-$\epsilon$ & -2.6$\epsilon$ & -3.6$\epsilon$ & -4.1$\epsilon$ & -4.3$\epsilon$ & -4.3$\epsilon$&&&\\
$\epsilon$ & 3.0$\epsilon$ & 4.6$\epsilon$ & 5.4$\epsilon$ & 5.7$\epsilon$ & 5.8$\epsilon$ & 5.8$\epsilon$&&\\
-$\epsilon$ & -2.7$\epsilon$ & -4.4$\epsilon$ & -5.5$\epsilon$ & -5.9$\epsilon$ & -6.1$\epsilon$ & -6.2$\epsilon$ & -6.2$\epsilon$&\\
...& ...& ...&...& ...& ...& ...& ...&
\end{tabular}
\end{center}
\caption{Round off error propagation with optimal step sequence in the BS extrapolation. \label{tab:roudoff2}}
\end{table}
Table~\ref{tab:roudoff2} shows the error propagation with the new step sequence.  We see that it performs much better than the original sequence suggested by \citet{Deuflhard1983}.

The second method to reduce the round off error in the extrapolation process is to extrapolate the raw increment $d\xi$ of the integration instead of the integrated result $\xi$. Since the GBS extrapolation is linear, it is possible to do the following
\begin{equation}
    \Delta T_{i,j} = \Delta T_{i, j-1} + \frac{\Delta T_{i,j-1} - \Delta T_{i-1, j-1}}{(n_i/n_{i-j+1})^2-1}.
\end{equation}
Because the raw increment $d\xi$ is usually tiny compared to the integrated results, the round of error in the first column of Table~\ref{tab:BS}  is $\sim {\rm eps}\times|d\xi|$ instead of $\sim {\rm eps}\times|\xi|$. Then, the round off error accumulated from the extrapolation table will be smaller than the error accumulated from the direct extrapolation. We note, however, that the initial $\Delta T_{i,1}$ should be the raw increment evaluated from the basic integrator, not from the subtraction of $T_{i,1}(H) - T_{i,1}(0)$. The close value subtraction will introduce additional round off errors and will significantly degrade the advantage of this method. 

\subsubsection{Algorithmic regularization in arbitrary precision integrations}
The GBS standalone is not efficient in solving extremely eccentric orbits and very close pair-wise encounters. Thus, in {\tt SpaceHub} we implement the algorithmic regularized arbitrary precision method AR-ABITS to deal with extremely eccentric orbits and very close encounters with arbitrary precision.

\begin{figure}
      \includegraphics[width=0.45\textwidth]{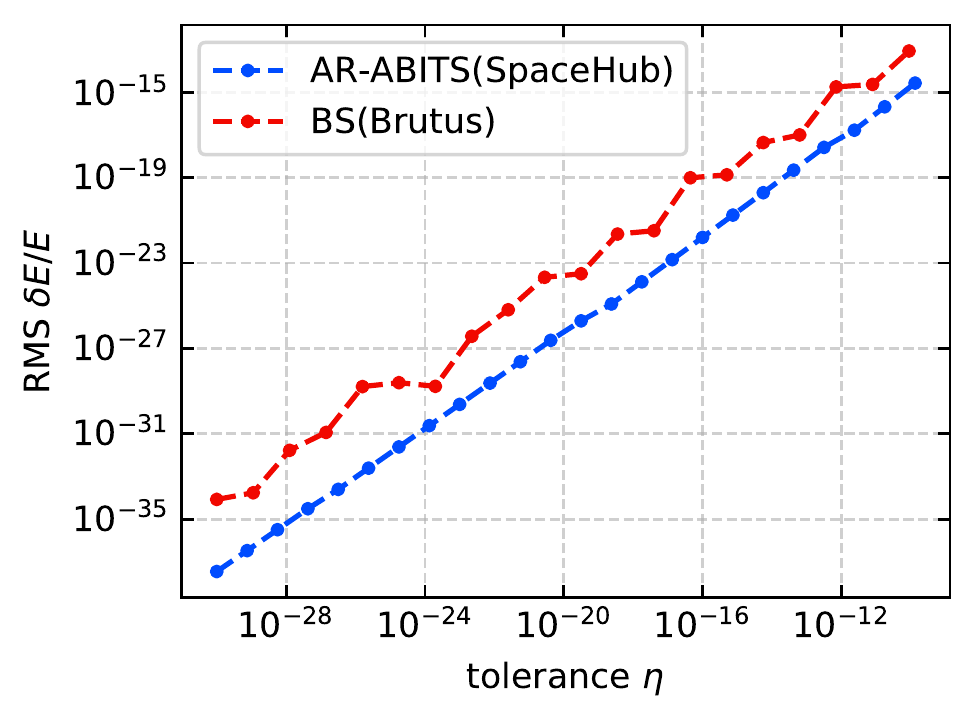}\\
       \includegraphics[width=0.45\textwidth]{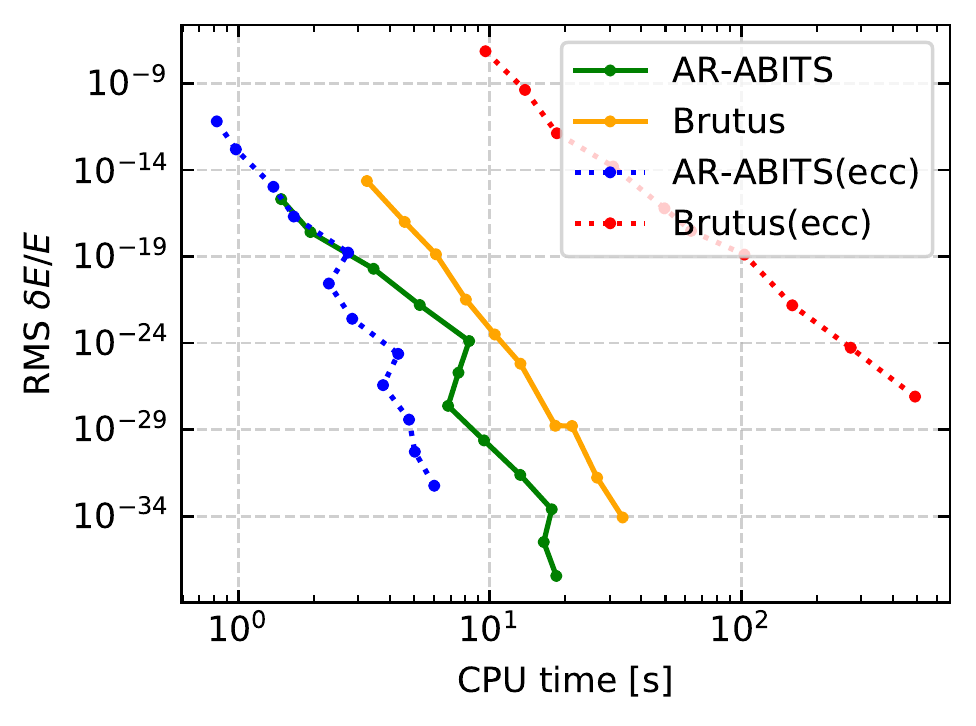}
    \caption{ Arbitrary precision methods in {\tt SpaceHub} and  {\tt Brutus}. {\it Upper panel:} Root mean square relative error as a function of the relative integration tolerance for integrations of the sun-earth-moon system for 100 moon orbits. {\it Bottom panel:} Root mean square relative error as a function of CPU time for integration of the sun-earth-moon system for 100 moon orbits (solid line), and for the integration of 100 orbits of an eccentric sun-earth two body system with e=0.9999 (dashed line). The mantissa bit is chosen from $4\times[\log_{10}(\eta)]_{\rm floor}+32$ as suggested by {\tt Brutus}. }
    \label{fig:arbitrary}
\end{figure}

 We find that the fine-tuned regularized GBS extrapolation works efficiently for arbitrary bits floating-point numbers.  The upper panel of Figure~\ref{fig:arbitrary} shows the relative energy error as a function of the relative tolerance $\eta$ for the GBS method in {\tt Brutus} and the AR-ABITS method in {\tt SpaceHub}. The integration is performed on a sun-earth-moon system for 100 moon orbits. For a relative tolerance $\eta$, $N = 4\times[\log_{10}(\eta)]_{\rm floor} + 32$ mantissa bits for floating-point numbers will be used as suggested by {\tt Brutus}. We can see from the upper panel that, for the same floating-point number bits and tolerance $\eta$, the BS extrapolation in {\tt SpaceHub} can achieve 1-2 orders of magnitude higher precision than {\tt Brutus}. With our regularization improvements and our special treatment for reducing round-off errors, we achieve even better precision in AR-ABITS. This trend becomes even more significant for low $\eta$, as round off errors become more problematic in higher order extrapolations.

The bottom panel of Figure~\ref{fig:arbitrary} shows the error scaling of {\tt Brutus} and AR-BITS. With the same choice of mantissa bits, we see that for a non-eccentric sun-earth-moon system, AR-BITS (green solid line) achieves the same precision but with a CPU time several times faster than in {\tt Brutus} (orange solid line). For eccentric systems with e=0.9999, AR-BITS (blue dotted line) achieves the same precision but with a CPU time one to two orders of magnitude faster than with {\tt Brutus} (red dotted line). To conclude, we find that AR-BITS is much more bit/time efficient in both eccentric and circular systems.

Our improvement on the GBS extrapolation can be applied to any algorithm based on GBS extrapolation, and is ideal for integrations requiring high precision with lower round off errors.

\subsection{AR-chain$+$: Improved AR-chain}\label{sec:archain+}
In the following, we will describe the improvements of AR-chain$+$ which make it superior to AR-chain.

\subsubsection{Improvement on the chain coordinate transformation}
In the original chain coordinates transformation, for system with N Cartesian coordinates, the transformed chain coordinates has N-1 coordinates as described in Equation~\ref{eq:chain}. In the inverse transformation, the centre of mass reduction is necessary when calculating the kinetic energy $T =  \sum_i \frac{1}{2}m_i \mathbf{v}_i^2$, which is required by the regularization method. For direct summation N-body integration in the large-N regime, most of the CPU time is used to evaluate the acceleration, since the cost of acceleration evaluation scales with $O$(N$^2$). Thus, the cost of centre of mass reduction that scales with $O$(N) is negligible. However, in the small-N regime, the cost of the centre of mass reduction can be significant.

Unlike the original transformation between $(\mathbf{x}_1,...,\mathbf{x}_N)$ $\leftrightarrow$ $(\mathbf{X}_1,...,\mathbf{X}_{N-1})$, we propose a new modified transformation such that the centre of mass reduction can be eliminated. Indeed, a bijective mapping between $(\mathbf{x}_1,...,\mathbf{x}_N)$ $\leftrightarrow$ $(\mathbf{X}_1,...,\mathbf{X}_N)$ can be constructed, where
\begin{eqnarray}
 \mathbf{X}_k &=& \mathbf{x}_{i_{k+1}}-\mathbf{x}_{i_k} \quad k = 1,2,...,N-1\\
 \mathbf{X}_N &=& \mathbf{x}_{i_1}\\
 \mathbf{V}_k &=& \mathbf{v}_{i_{k+1}}-\mathbf{v}_{i_k} \quad k = 1,2,...,N-1\\
 \mathbf{V}_N &=& \mathbf{v}_{i_1}
\end{eqnarray}
and
\begin{eqnarray}
 {\mathbf{x}}_{i_1} &=& \mathbf{X}_N\\
 {\mathbf{x}}_{i_{k+1}} &=& {\mathbf{x}}_{k} + \mathbf{X}_k \quad k = 1,2,...,N-1\\
 {\mathbf{v}}_{i_1} &=& \mathbf{V}_N\\
 {\mathbf{v}}_{i_{k+1}} &=& {\mathbf{v}}_{k} + \mathbf{V}_k \quad k = 1,2,...,N-1
\end{eqnarray}
with equations of motion
\begin{eqnarray}
  \frac{d\mathbf{X}_k}{dt} &=& \mathbf{V}_k,\quad\quad k=1,2,...,N\\
    \frac{d\mathbf{V}_k}{dt} &=& \mathbf{g}_{i_{k+1}}- \mathbf{g}_{i_k} + \mathbf{f}_{i_{k+1}}-\mathbf{f}_{i_{k}} + \mathbf{f}_{i_{k+1},v}-\mathbf{f}_{{i_k},v}\\
    k&=&1,2,...,N-1\\
    \frac{d\mathbf{V}_N}{dt} &=& \mathbf{g}_{i_1} + \mathbf{f}_{i_1} + \mathbf{f}_{i_1,v}\,.
\end{eqnarray}
This transformation preserves the centre of mass reduction without the need for any additional acceleration evaluation.

\begin{figure}
      \includegraphics[width=.45\textwidth]{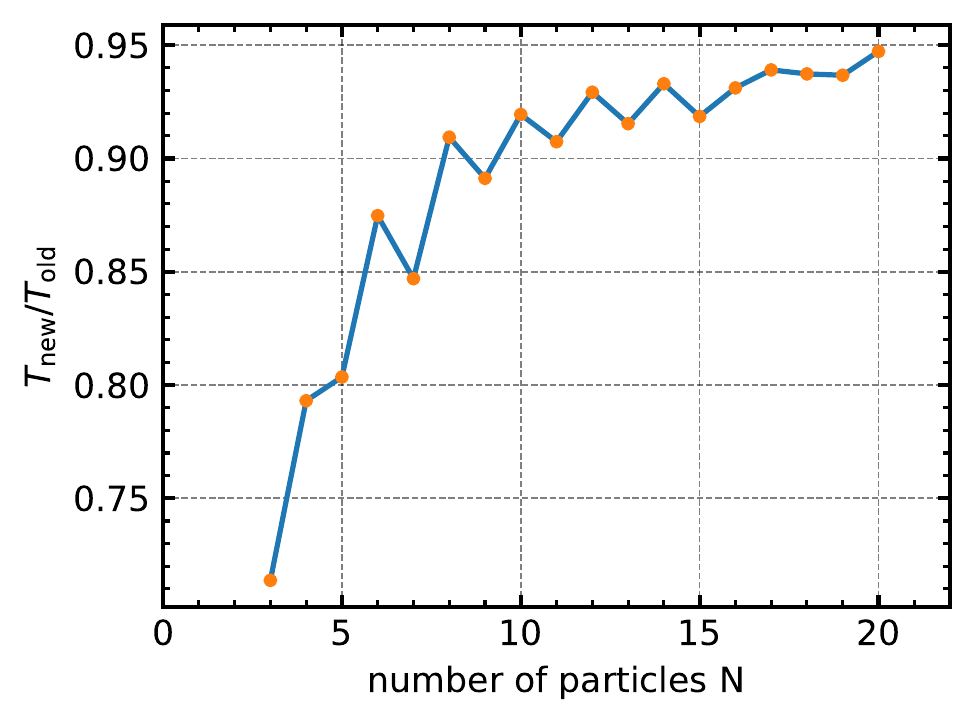}
    \caption{The CPU time of the improved chain transformation $T_{\rm new}$ versus the old chain transformation as a function of the particle number N.}
    \label{fig:chain-improve}
\end{figure}

Figure~\ref{fig:chain-improve} shows the relative CPU time cost of this new transformation compared to the old transformation, as a function of the particle number N. In the small-N regime, the new transformation that eliminates the centre of mass reduction can save significantly on computational run time.  In the large N-regime, since the acceleration evaluation takes up most of the CPU time, the two transformations are almost identical in CPU cost. 

\subsubsection{Introducing the active round off error compensation into AR-chain}\label{sec:err redu}
Other than the round off error reduction in GBS extrapolation, the active round off error reduction can also be used in the AR-chain algorithm. 

 Instead of implementing the active compensation process during a certain step of the algorithm as described in Equation~\ref{eq:active} and avoid writing the compensation procedure everywhere, we implement several new floating point number types that execute the active compensation automatically. Specifically, three additional floating-point number types in both 32 and 64 bit, namely Kahan number($\rm float\_k$, $\rm double\_k$), Neumaier number($\rm float\_p$, $\rm double\_p$) and Klein number($\rm float\_e$, $\rm double\_e$), with different active error compensation strategies are implemented in {\tt SpaceHub} as base floating point number types. In those floating-point number types, the operator '+=' can automatically perform the active compensation process. Therefore, any algorithm in {\tt SpaceHub} taking these floating point data types can directly activate the active round off error compensation.  The cost for this implementation is a higher cache latency and double sized memory to save the error from last addition. However, in the small-N regime, these costs are not significant.
 
Usually, the active round off error compensation can reduce the round off error by 1-2 order of magnitude. However, we found that in AR-chain, the effect is significantly degraded by the chain coordinates.  After each step, to make sure that the shortest relative position is kept in the chain coordinates,  the chain coordinates may need to be updated. Once the chain coordinates update is required, then the chain coordinates $X_{k,{\rm new}}$ need to be reconstructed from $X_{k,{\rm old}}$. Therefore, the round off error ${\rm err}_{+}(X_{n}, dX_{n})$ calculated from the last step, which is computed for $X_{k,{\rm old}}$, becomes incorrect if applied to $X_{k,{\rm new}}$. Thus, if the chain update is frequent during the integration, the active round off error compensation will be interrupted and then it becomes less useful. The branching coordinates proposed by \citet{rantala_mstar_2020}, that need fewer coordinates updates has the potential to solve this problem. We will investigate this in the future.

\begin{figure}
      \includegraphics[width=.49\textwidth]{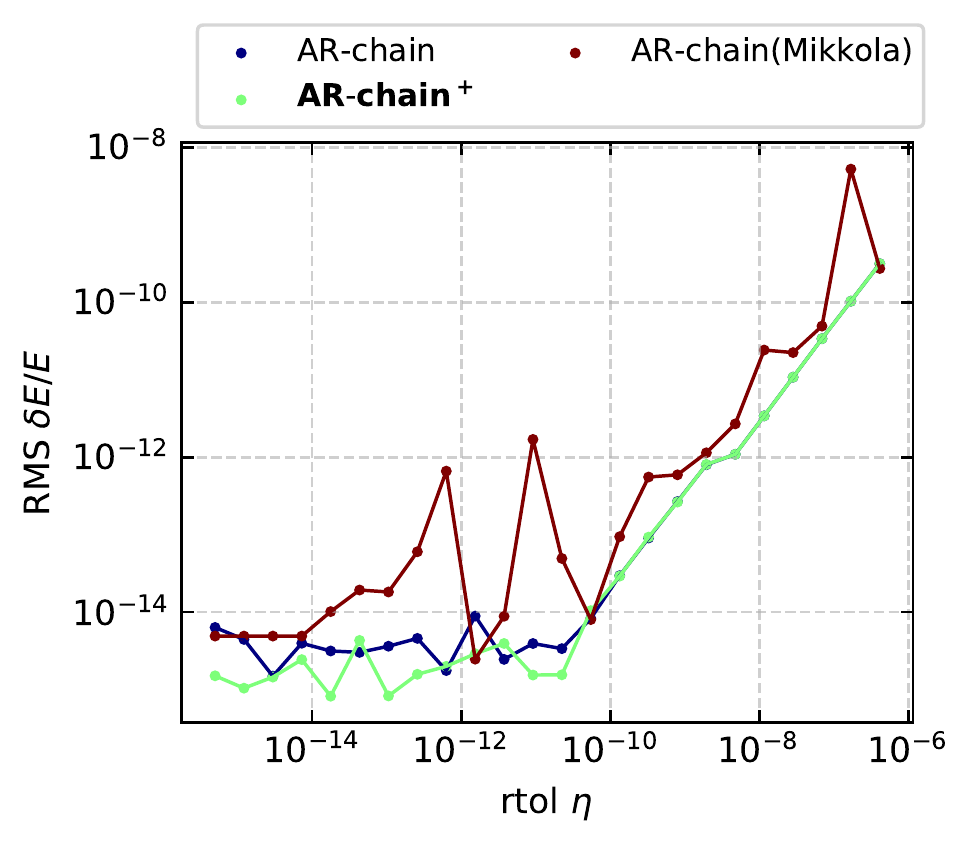}
    \caption{Integration on the sun-earth-moon system with 1000 moon orbits. 5000 equal spaced relative energy error during the 1000 orbits are outputted to calculate the root mean square relative energy error. This figure shows the rms relative energy error as a function of the relative tolerance $\eta$ for Mikkola's AR-chain, AR-chain in {\tt SpaceHub} and AR-Chain$^+$.}
    \label{fig:ARC+}
\end{figure}

We integrate the sun-earth-moon system with 1000 moon orbits with the original Mikkola's AR-chain, AR-chain in {\tt SpaceHub} and AR-Chain$^+$ with different relative error tolerance $\eta$. Figure~\ref{fig:ARC+} shows the rms relative energy error as a function of $\eta$.

\subsection{AR-Sym6+: Algorithmic regularized higher order symplectic method}\label{sec:arsym6+}

The original AR-chain uses the leapfrog scheme as the basic algorithm. Due to its time symmetry, it does very well in conserving the total energy of the Hamiltonian system. However, the leapfrog scheme alone is just a second order method that, when used alone, is rarely sufficiently accurate. Therefore, to achieve higher precision, AR-chain adopts the GBS extrapolation scheme which can efficiently construct higher order methods from lower order methods.  However, the round-off errors from the BS method can be significant due to the extrapolation process. Several measures can be taken to reduce them, but the reduction can be limited if higher order methods are included. Therefore, theoretically, even BS extrapolation can achieve arbitrarily high order accuracy.  However, we find that, for most cases of interest, 13th to 15th order works best for double precision floating point numbers. Higher order extrapolations will be capped by the round off error. More importantly, the extrapolation can break the time symmetry of the leapfrog scheme, and can significantly affect the relative energy error in the round off error regime. For integration methods based on the BS extrapolation with double precision FP numbers as the arithmetic type, the relative energy error for short time integration is usually at the level of $10^{-13}$ due to the extrapolation. As discussed before, the active error compensation can reduce the error further to $\sim 10^{-14}$, but it becomes hard to achieve higher precision under the BS extrapolation scheme.

Using higher order symplectic methods to replace the BS extrapolation has been mentioned in \citet{mikkola_implementing_2008}. However, to date, there have been no implementations of higher order symplectic methods with algorithmic regularization. Benefiting from the code architecture of {\tt SpaceHub}, we can easily glue different algorithms together, and introduce the regularization and chain algorithm into the higher order symplectic method. Without the rational extrapolation, the time symmetry of the symplectic method can be preserved.  Thus, this new method can efficiently deal with extreme eccentricity systems and very close pair-wise approaches with even higher precision. More importantly, with regularization, it is possible to use fixed step sizes to solve the time evolution of extremely eccentric systems. The fixed step size regularized method maintains the time symmetry while avoiding these issues arising from the extrapolation. Consequently, we can achieve higher precision in the round off error regime (i.e., when round off errors dominate the growth of the total energy error budget). 

\subsubsection{Error estimation}
For an integration method of order $k$, the numerical integration $I_n(f)$ on $f$ for total step length $H$ with $n$ intermediate steps has an error that scales as
\begin{equation}
    E_n(f) = |I_n(f) - F| \sim C \left(\frac{H}{n}\right)^{k}\,,
\end{equation}
where $F$ is the unknown true value of the integration and $C$ is a constant. To estimate $E_n(f)$, we can do additional integrations on $f$ with $m$($>n$) steps. The error for $I_{m}(f)$ is
\begin{equation}
    E_{m}(f) = |I_{m}(f)-F| \sim C \left(\frac{H}{m}\right)^{k} = \frac{E_{n}(f)}{(m/n)^k}.
\end{equation}
Then we can estimate the error of the integration $I_{m}(f)$,
\begin{equation}
   E_{m} \sim \frac{1}{(m/n)^k-1}|I_{m} -I_{n}|\,.
\end{equation}

To obtain a converged result with tolerance $\epsilon$, a sequence of step numbers $n_1, n_2,...,n_j$ can be performed such that the dimensionless error is less than unity,
\begin{equation}\label{eq:err}
    {\rm error} = \frac{|I_{n_j}-I_{n_{j-1}}|}{[(n_j/n_{j-1})^k-1]|I_{n_k}|\epsilon} < 1\,.
\end{equation}
From our experiments, we find that the choice of $n_j$ for efficient convergence and speed works well for $n_j=2j$ with $j=1,2,...$.

\subsubsection{Step size control}
As discussed above, for an integration method of order $k$, the global truncation error scales as $\sim H^k$, where $H$ is the step size. For integrations working in the asymptotic regime, we have 
\begin{equation}
    \frac{H_{\rm old}^k}{\rm error_{\rm old}} \sim \frac{H_{\rm new}^k}{\rm error_{\rm new}}
\end{equation}
To get the converged result in the next step, i.e $\rm error_{\rm new} < 1$, the simplest way to chose the step size in the next step would be just
\begin{equation}
    H_{\rm new} = H_{\rm old} (\frac{1}{\rm error_{\rm old}})^{1/k}\,.
\end{equation}
However, the integration would not always work effectively in the asymptotic regime especially, but also in the round off error regime, where the round off error becomes non-negligible compared to the truncation error. Thus, more complicated error/step control might be needed.

In control theory, the Proportional–Integral–Derivative (PID) controller is believed to be efficient and useful in performing continuous error controlling. In {\tt SpaceHub}, the PI step size controller that applies proportional and integral feedback to the integration system is the default step size controller, where the new step will be given by
\begin{equation}
    H_{\rm new} = S_1 \left(\frac{S_2}{\rm error}\right)^{C_P/k}\left(\frac{\rm error}{S_2}\right)^{C_I/k}H_{\rm old}\,,
\end{equation}
where $C_P$, $C_I$ are, respectively, the coefficients of proportionality and the integral feedback part. The specific values of $C_P$ and $C_I$ are problem-dependent but a general value could be $C_P =0.7$ and $C_I=0.4$ \citep{nr2002}. The variables $S_1$ and $S_2$ are the safe factors, which tend to be few percents smaller than one to make sure the new step size has higher probability to get a converged result.  To avoid varying the step size rapidly, a time step limiter is applied to the $H_{\rm new}$ calculation. The new step size is constrained within the range
\begin{equation}
    \frac{S_3^{1/k}}{S4} < \frac{H_{\rm new}}{H_{\rm old}} < \left(\frac{1}{S_3}\right)^{1/k}\,
\end{equation}
where $S_3=0.02$ and $S_4 =4$.

\subsubsection{Choosing the order of the Symplectic methods}
\citet{yoshida_construction_1990} gives an efficient way to construct higher order symplectic methods recursively using lower order symplectic methods via the Baker-Campbell-Hausdorff formula. From \citet{yoshida_construction_1990}, if a symmetric integrator of order 2$n$, $S_{2n}(\tau)$ is already known, a $2n+2$ th order integrator can be obtained via the product
\begin{equation}
S_{2n+2} =S_{2n}(z_1\tau)S_{2n}(z_0\tau)S_{2n}(z_1\tau)
\end{equation}
where $z_0$ and $z_1$ satisfy
\begin{equation}
    z_0+2z_1=1, z_0^{2n+1}+2z_1^{2n+1}=0\,.
\end{equation}
Then, starting from the 2nd order leapfrog scheme with $S_{2}(\tau) = D(\tau/2)K(\tau)D(\tau/2)$, higher order symplectic integrators can be constructed recursively. In {\tt SpaceHub}, we implement symplectic methods up to 10th order. 

For symplectic integrators using the regularization chain method, we test from 6th order to 10th order. From our experiments, we find that the 6th order method (hereafter called AR-sym6) works best in terms of precision with decent convergence speed for 64 bits FP data type. Using the higher order symplectic methods directly without extrapolation preserves the time symmetry of the integration.  Thus, it achieves much higher precision relative to the original AR-chain, using the active round off error compensation.

\begin{figure}
      \includegraphics[width=.49\textwidth]{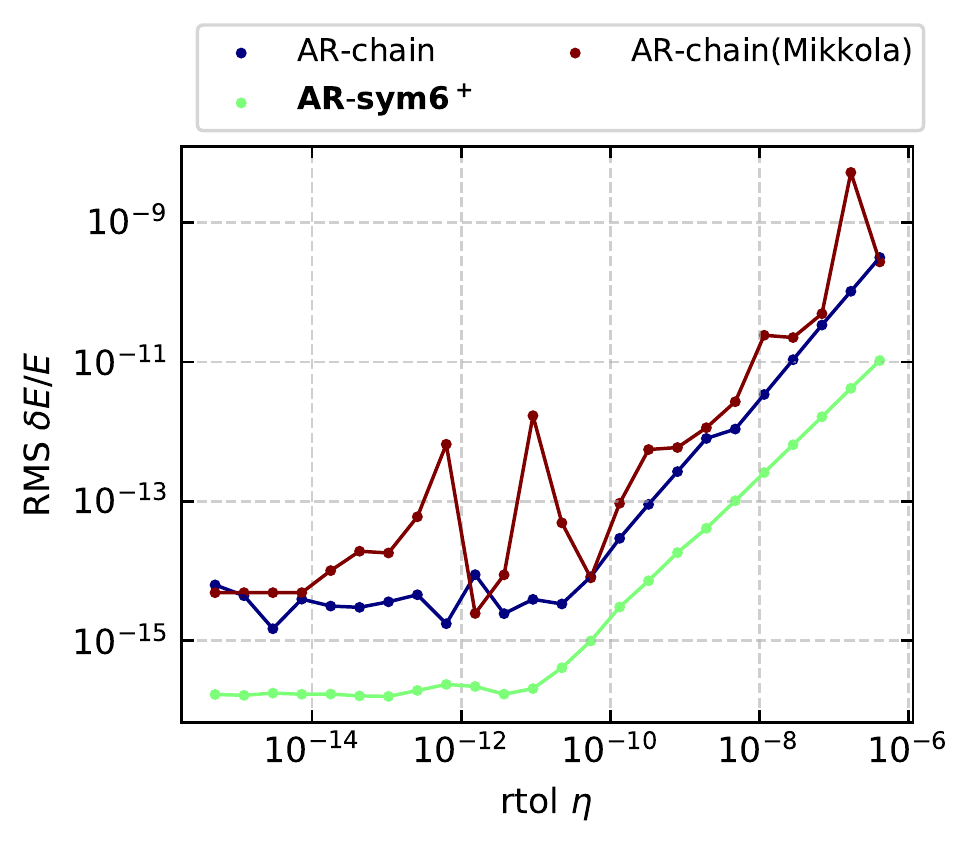}
    \caption{Same as Figure~\ref{fig:ARC+}, but testing the AR-sym6$^+$.}
    \label{fig:ARsym6}
\end{figure}

We integrate the sun-earth-moon system with 1000 moon orbits with the original Mikkola's AR-chain, AR-chain in {\tt SpaceHub} and AR-Sym6$^+$ with different relative error tolerance $\eta$. Figure~\ref{fig:ARsym6} shows the rms relative energy error as a function of $\eta$.

\subsection{AR-Radau+: introducing algorithmic regularization into Gauss-Radau integration}\label{sec:arradau+}
Gauss-Radau spacings are useful in solving few-body problems accurately, especially after the active round off error compensation was implemented in {\tt Rebound}. However, the normal Gauss-Radau method is not efficient/accurate enough to solve extremely eccentric orbits as well as close encounters because tiny steps are required to solve the trajectories at the point of closest approach. To solve this problem while keeping the advantages of the Gauss-Radau stepping, we introduce the algorithmic regularization into the Gauss-Radau method.

The equations of motion of the regularized system are first order differential equations.  For a system with Hamiltonian $H$, the LogH method gives the equations of motion as,
\begin{eqnarray}
dt/dh &=& 1/(T+B)\\
d\mathbf{r}/dh &=& \mathbf{v}/(T+B)\\
d\mathbf{v}/dh &=& \mathbf{a}/U\\
dB/dh &=& \partial H/\partial t/U\,,
\end{eqnarray}
while the TTL method gives the equations of motion as,    
\begin{eqnarray}
dt/dh &=& 1/\omega\\
d\mathbf{r}/dh &=& \mathbf{v}/\omega\\
d\mathbf{v}/dh &=& \mathbf{a}/\Omega\\
d\omega/dh &=& \partial \Omega/\partial t/\Omega\,.
\end{eqnarray}
They work smoothly with the leap-frog based method because advancing the velocity and position are divided into 'kick' and 'drift' separately. To solve the equations of motion of the regularized system with the Gauss-Radau method, a general coordinate is required where,
\begin{equation}
    \mathbf{y}=(\mathbf{r},\mathbf{v},\omega,B,t)
\end{equation}
such that $\mathbf{y}|_h$ can be expanded into 
\begin{equation}\label{eq:ar-radau}
    \frac{d\mathbf{y}}{dh}\bigg|_h \sim \frac{d\mathbf{y}}{dh}\bigg|_0 + \mathbf{b_1}h+\mathbf{b_2}h^2+...+\mathbf{b_7}h^7
\end{equation}
where 
\begin{equation}
    \frac{d\mathbf{y}}{dh}\bigg|_0 = (\frac{d \mathbf{r}}{dh}, \frac{d\mathbf{v}}{dh},\frac{d\omega}{dh},\frac{dB}{dh},\frac{dt}{dh})
\end{equation}
The same process as Equation~\ref{eq:radau-b}  and \ref{eq:radau-g} can be performed to calculate the $\mathbf{b}_i$. Then the general coordinates at the end of the step $dt$ can be obtained by analytically integrating Equation~\ref{eq:ar-radau}.
\begin{equation}
    \mathbf{y}|_h \sim \mathbf{y}|_0 + hdt\bigg(\frac{d\mathbf{y}}{dt}\bigg|_0 +\frac{h}{2}\bigg(\mathbf{b_1}+\frac{2h}{3}\bigg(\mathbf{b_2}+...\bigg)\bigg)\bigg)\\
\end{equation}

The chain algorithm can also be constructed by replacing the equations of motion for $\mathbf{r}$ and $\mathbf{v}$ with equations of motion for $\mathbf{X}$ and $\mathbf{V}$, as described in Equation~\ref{eq:chain}.

\begin{figure}
      \includegraphics[width=.49\textwidth]{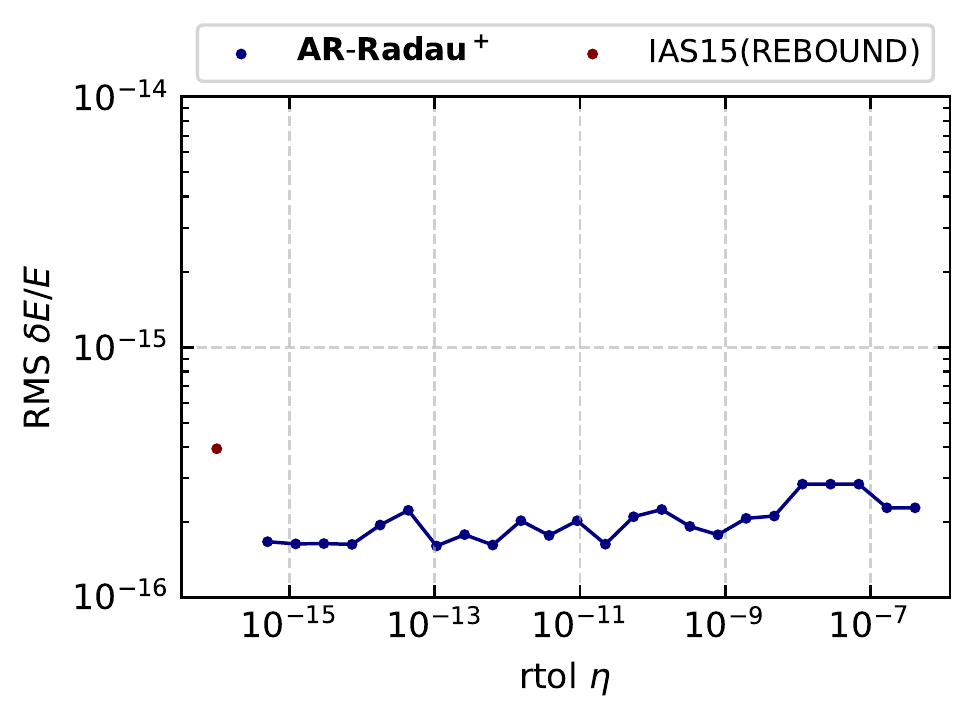}\\
      \includegraphics[width=.49\textwidth]{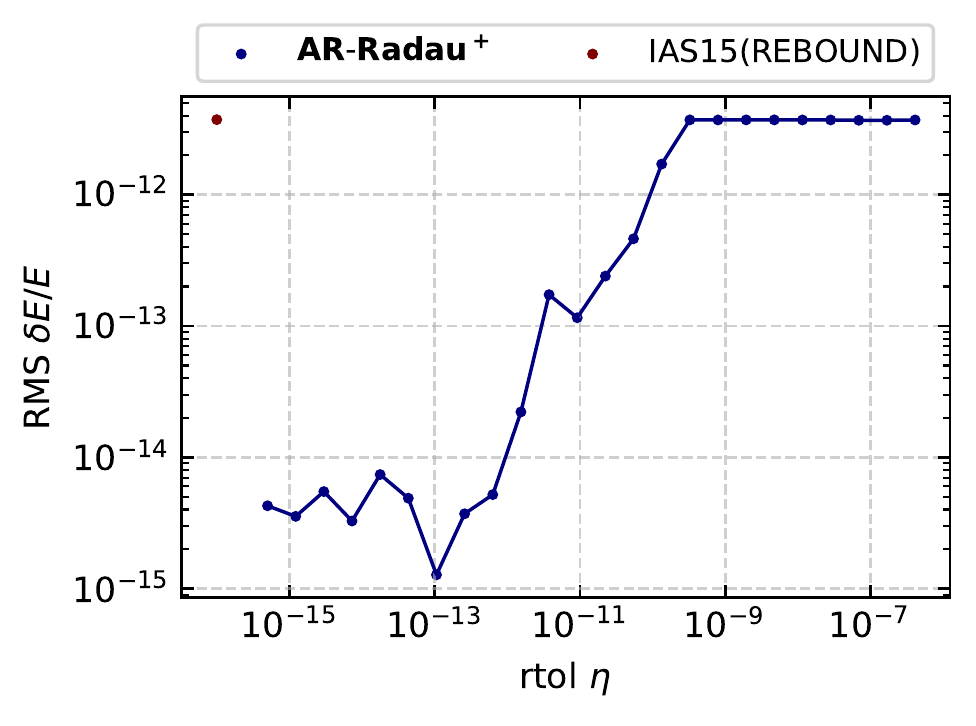}
    \caption{{\it Upper panel}: Same test as in Figure~\ref{fig:ARC+} for IAS15 and AR-Radau$^+$. {\it Bottom panel}: Test of 1000 orbits on the two body eccentric system with a = 1 AU and e = 0.9999 consists of $M_1 = 1 M_\odot$ and $M_2 = 1 M_\oplus$. The IAS15 does not provide API to change the $\eta$, thus the $\eta$ is fixed as default $10^{-16}$. }
    \label{fig:ARRadau}
\end{figure}

Similar to previous subsections, we integrate two systems with different relative error tolerance $\eta$. The first test integrates the sun-earth-moon system with 1000 moon orbits and the second test integrates an eccentric two body orbit with $M_1 = 1 M_\odot$, $M_2 = 1 M_\oplus$, a = 1 AU and e = 0.9999 with 1000 orbits. The IAS15 does not provide API to change the $\eta$, thus the $\eta$ is fixed at the default value $10^{-16}$.  

The upper panel of Figure~\ref{fig:ARsym6} shows the rms relative energy error as a function of $\eta$ for the IAS15 and AR-Radau$^+$ methods for the sun-earth-moon system.  The bottom panel shows the same results for the eccentric two body system.

\section{Performance tests for real astrophysical systems}
In this section we test the precision and performance of the integration methods adopted in {\tt SpaceHub}, and compare them to other high-precision few-body codes, including {\tt Brutus}, {\tt Rebound}, {\tt ABIE}\footnote{ABIE is a new GPU-accelerated direct N-body code. The integrator adopted in these simulations is a 15th-order Gauss-Radau algorithm with an adaptive timestep scheme. The algorithm is particularly optimized for close encounters.}, Mikkola's {\tt AR-chain}, and so on. 

All tests are performed on an Intel-i7-8700k CPU under a Linux OS with GCC-10.2.0. All codes are compiled with the {\tt -O3} compile optimization option. Table~\ref{tab:tested-methods} shows the algorithms tested in each section.

\begin{table}
\begin{center}{}
\begin{tabular}{ |l|l|l|l|l |} 
\hline
& algorithm          & code     & setups     & link             \\
\hline
 & Bulirsch-Stoer            & {\tt SpaceHub}       &  rtol = 10$^{-14}$, atol = 0   &  \href{https://github.com/YihanWangAstro/SpaceHub}{source code}          \\
 & AR-Chain            & {\tt SpaceHub}       &   rtol = 10$^{-14}$, atol = 0     &  \href{https://github.com/YihanWangAstro/SpaceHub}{source code}       \\
 & AR-Chain$^+$            & {\tt SpaceHub}       &  rtol = 10$^{-14}$, atol = 0  &  \href{https://github.com/YihanWangAstro/SpaceHub}{source code}      \\
 & AR-Radau$^+$            & {\tt SpaceHub}       &    code fixed default & \href{https://github.com/YihanWangAstro/SpaceHub}{source code}           \\
 & AR-sym6$^+$            & {\tt SpaceHub}       &   rtol = 10$^{-14}$, atol = 0    &  \href{https://github.com/YihanWangAstro/SpaceHub}{source code}        \\
  & AR-ABITS            & {\tt SpaceHub}       &   rtol = 10$^{-14}$, atol = 0    &  \href{https://github.com/YihanWangAstro/SpaceHub}{source code}        \\
 & IAS15            & {\tt Rebound}     &  code fixed default&  \href{https://github.com/hannorein/rebound}{source code}           \\
 & Radau            & {\tt ABIE}       & code fixed default  &  \href{https://github.com/MovingPlanetsAround/ABIE}{source code}       \\
 & Bulirsch-Stoer            & {\tt Brutus}       &  rtol = 10$^{-14}$, atol = 0&   \href{https://github.com/amusecode/amuse/tree/master/src/amuse/community/brutus}{source code}     \\
 & AR-chain            & {\tt Mikkola}       &  rtol = 10$^{-14}$, atol = 0 & \href{http://www.astro.utu.fi/mikkola/}{source code}      \\
\hline
\end{tabular}
\end{center}
\caption{Tested algorithms\label{tab:tested-methods}}
\end{table}

\subsection{Precision \& Performance Tests}
For the precision and performance tests, only the Newtonian interactions will be included. The precision indicator, specifically the relative energy error, will be evaluated as
\begin{equation}
    \frac{\delta E}{E}(t) = \frac{|E(t)-E(0)|}{|E(0)|}
\end{equation}
with $E = T-U$.  For algorithms with regularization, to decrease the round-off errors from the potential energy calculations, the estimation becomes\citep{mikkola_implementing_2008}
\begin{equation}
    |\log(\frac{T+B}{U})| = |\log(1+\frac{\delta E}{U})|\sim \frac{|\delta E|}{U},
\end{equation}
where $B$ is defined in Equation~\ref{eq:b} as the binding energy that will be evaluated at $t=0$ and will continue to evolve in time with the regularized system. For each test, 5000 equally spaced $\frac{\delta E}{E}(t)$ computations will be outputted during the integration and the root mean square (RMS) error is calculated as
\begin{equation}
    {\rm RMS}\quad \frac{\delta E}{E} = \sum_i^N \sqrt{ (\frac{\delta E}{E}(t_i))^2/N }\,.
\end{equation}

For our performance tests, the CPU wall time is measured from the best performance test of 5 repeat runs with the same initial conditions.  This is done to avoid CPU interruption by other processes operating in the background. The wall time only quantifies the time evolution within the main loop of the integration scheme, and ignores all initialization and finalization procedures in order to obtain normalized comparisons between the different algorithms. The IOs need non-negligible CPU time in short term performance tests. To precisely measure the wall time of different algorithms in our performance tests, all outputs are turned off.  The test file and initial conditions for our test cases can be found at \href{https://github.com/YihanWangAstro/SpaceHub/tree/master/test/regression_test}{SpaceHub-Tests}. Tests of other codes are performed using the code links provided above with the same initial conditions.

Since every integration method can be well-tuned for a specific problem, it can be tricky to do proper comparisons between different integration methods. However, for each integration method included in {\tt SpaceHub}, there are always default parameters, the so called 'out-of-the-box' parameters, which help control the integration flow and which vary among different integration schemes. To make the comparisons as normalized as possible, we use the out-of-the-box parameters for all integration methods when performing all comparison tests in order to ensure a fair comparison between the different methods adopted in different codes (e.g., \texttt{REBOUND}, \texttt{ABIE}, etc.). One should also note that the basic math functions like 'pow', 'sin', 'cos' etc. in the standard math library are not platform-independent. Thus, algorithm (including \texttt{ABIE}, \texttt{Mikkola}'s AR-chain and \texttt{Brutus}) that use those math functions become platform-independent as well. The test results could be slightly different on different platforms. A platform-independent math library will be implemented in the next version of the {\tt SpaceHub}.   

\subsubsection{The Earth-Moon-Sun System}

We now pay attention to a simple sun-earth-moon system that is easy to integrate for every method. The integrated system consists of the sun, the earth and the moon. We integrate the system for 1000 moon orbits using adaptive time stepping, and quantify the performance of each method.

\begin{figure*}
      \includegraphics[width=\textwidth]{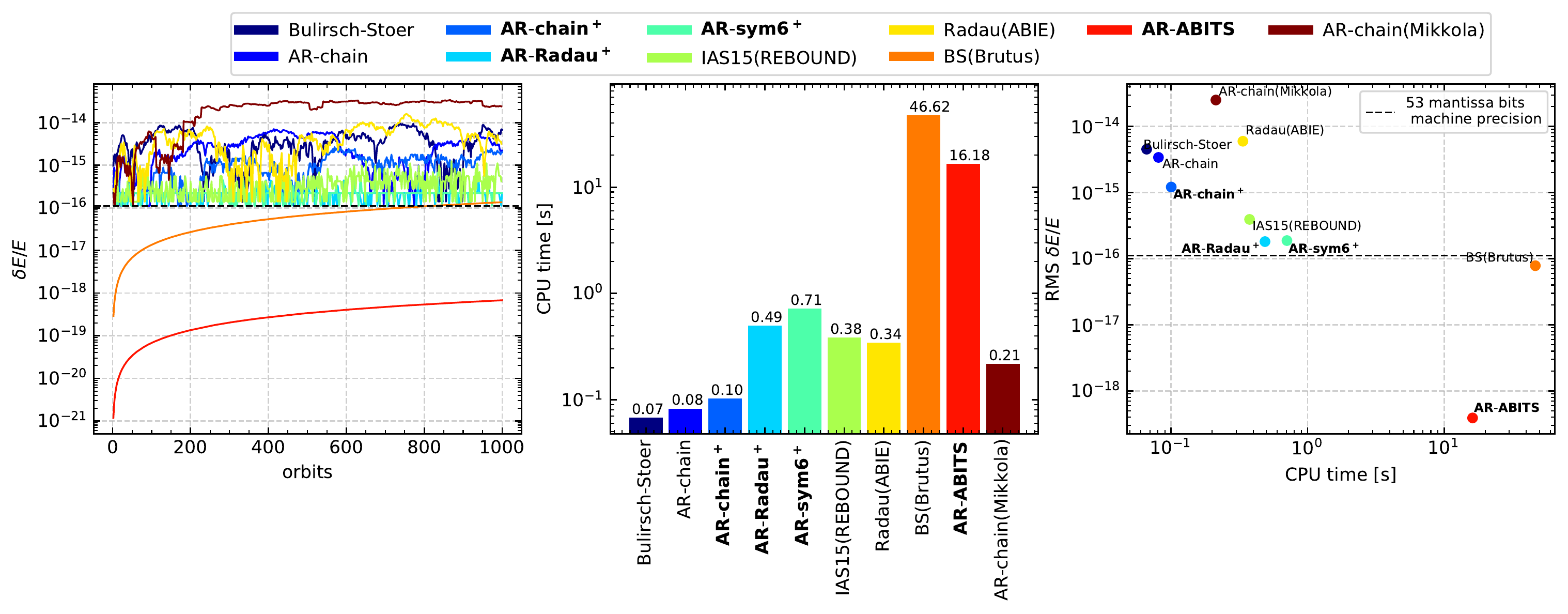}
    \caption{Relative energy error and performance tests on the {\bf sun-earth-moon system}. The integration duration is 1000 moon orbits. {\it Left panel:} Relative energy errors for different integration methods in {\tt SpaceHub} (without parentheses) as a function of the number of orbital periods and for different integration methods adopted in other codes (with the names of the codes indicated in parentheses). {\it Middle panel:} CPU wall time for each integration method. {\it Right Panel:} Root mean square relative energy error (accumulated over time) versus CPU wall time. The relative tolerance is $10^{-14}$ and the absolute tolerance is $0$ for methods that can be adjusted. All methods use IEEE-754 double precision floating point numbers, except for 'AR-ABITS' and 'BS(Brutus)'. 'AR-ABITS' and 'BS(Brutus)' use 88 mantissa bits and non-standard extended floating-point numbers. Methods with bold-faced font are new unique methods in {\tt SpaceHub}. The initial conditions and performance test descriptions can be found at \url{https://github.com/YihanWangAstro/SpaceHub/tree/master/test/regression_test}.}
    \label{fig:earth}
\end{figure*}

Figure~\ref{fig:earth} shows the precision and CPU wall time for each method. The left panel shows the relative energy error of each method as a function of time, the middle panel shows the CPU wall time for each method and the right panel shows the rms relative energy error versus CPU time. From the figure, we can see that the implementation of AR-chain in {\tt SpaceHub} is $\sim 2-2.5$ times faster than Mikkola's implementation. The AR-Chain$^+$ is slightly slower but comes along with with slightly higher precision. The Radau method in ABIE reaches the same precision but significantly slower than the AR-Chain-based methods in {\tt SpaceHub}. The IAS15, a Radau method with improvements implemented by \citet{rein_ias15_2014}, achieves higher precision near the machine precision with the same speed as the Radau method in ABIE. The algorithmic regularized Radau method in {\tt SpaceHub} and the algorithmic regularized 6th symplectic methods have better round off error control for this problem, and thus yield even higher precision computations than IAS15. The algorithmic regularized 6th order symplectic method is significantly slower than the other two due to its extra integration for error evaluation at each step. The AR-Radau method, with extended double sized coordinates and extra regularization function evaluation (which mainly spends its time on evaluating U) at each time step is only slightly slower than IAS15 for this test case. For arbitrary precision methods with non-standard floating-point types, we see that the algorithmic regularized arbitrary bits (AR-ABITS) method is faster and far more accurate than {\tt Brutus}, as previously discussed.

\subsubsection{Extremely eccentric systems}

For our second test case, we analyze a two body system with extreme eccentricity e=0.9999 and semi-major axis a = 1 AU. The central object has a mass of 1 $M_\odot$ and the test particle has a mass of 1 $M_\oplus$. We integrate the system for 1000 orbits. This test is designed to quantify how the different integration methods are able to handle extremely eccentric orbits and very close pair-wise approaches between particles.
\begin{figure*}
      \includegraphics[width=\textwidth]{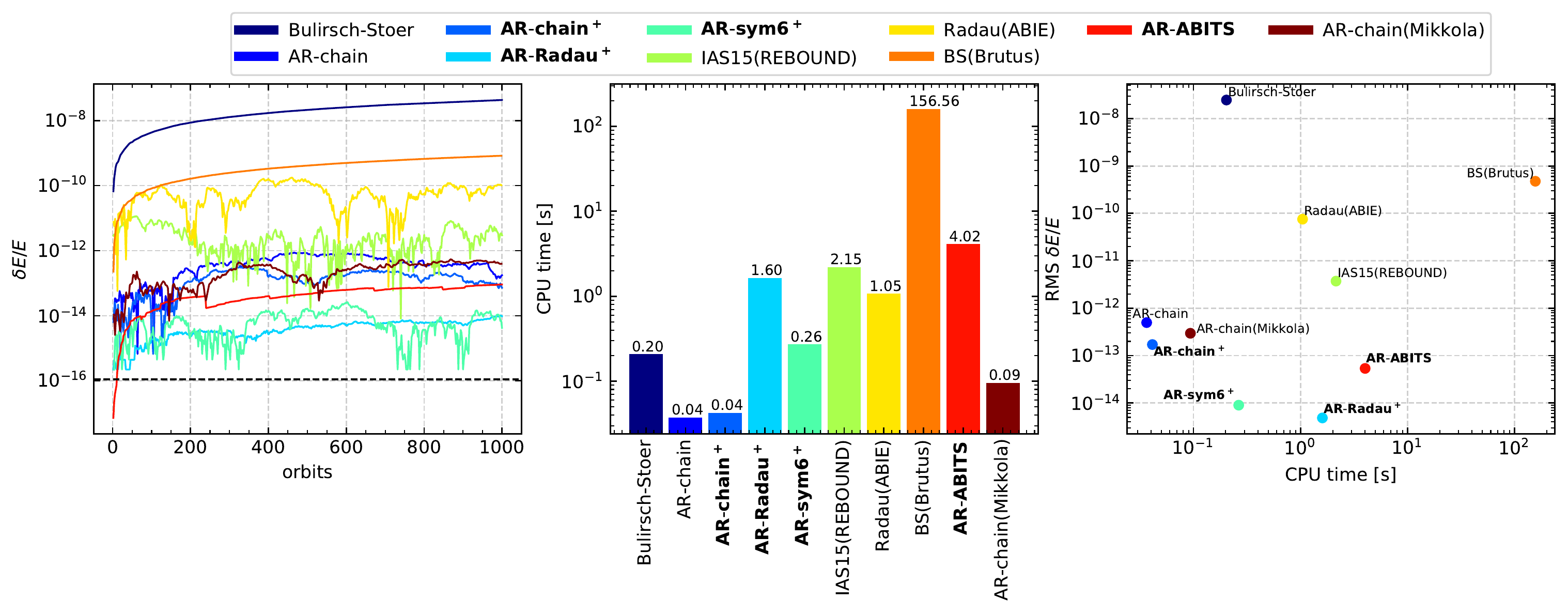}
      
    \caption{Similar test to Figure~\ref{fig:earth} but on an {\bf extremely eccentric two body system}. The central body has a mass of 1~$M_\odot$ and the orbiting body has a mass of 1~$M_\oplus$ with an initial semi-major axis of 1 au and an initial eccentricity of 0.9999. The integration duration is 1000 orbits.}
    \label{fig:ecc}
\end{figure*}

As in Figure~\ref{fig:earth}, we can see from  Figure~\ref{fig:ecc} that this system becomes challenging to integrate for methods without regularization. We find that the BS method accumulates errors quickly, rapidly reaching up to $\sim 10^{-8}$. The Radau method in ABIE can steadily maintain the error at the level of $\sim 10^{-10}$ for this integration duration. IAS15 behaves the best as a non-regularized method, keeping the error near $\sim 10^{-12}$, which is near the limit for non-regularized methods $\epsilon/(1-e_{\rm max})\sim 10^{-16}/(1-0.9999)$. For methods with regularization, we find that all integration schemes maintain a precision below $10^{-13}$ for this problem over the course of the integration duration. As for the simple test on the sun-earth-moon system, the AR-chain in {\tt SpaceHub} is roughly two times faster than Mikkola's AR-chain. Unlike this previous test case, the  AR-chain$^+$ method has nearly the same precision as AR-chain.  For this test case, the AR-chain$^+$ method achieves one order of magnitude higher precision. This is because there is no chain update in the two body system, such that the active error compensation can precision compensate the error from the last step in order to correct the coordinates for the next step. For systems with frequent chain updates, where the chain coordinates change frequently, the active error compensation become less useful. The AR-Radau method achieves 3 orders higher precision than IAS15 for this problem with faster speed. The AR-sym6 method reaches an energy error of $\sim 10^{-14}$ roughly ten times faster than IAS15. AR-ABITS is therefore better at dealing with extreme eccentricities and close encounters, achieving 4 orders of magnitude higher precision with a run time that is roughly 40 times faster than {\tt Brutus} for this test problem.

This test displays the advantage of the regularization schemes in dealing with extreme eccentricities and close encounters. With regularization, we can always get better results with faster speed. In {\tt SpaceHub}, one can introduce the regularization algorithm in any method provided that it is used the 'regularized system' as the integration scheme.

\subsubsection{Outer Solar System}

Next, we test the outer solar system.  We consider four planets: Jupiter, Saturn, Uranus and Neptune and evolve the system for 1000 Jupiter orbits.

\begin{figure*}
      \includegraphics[width=\textwidth]{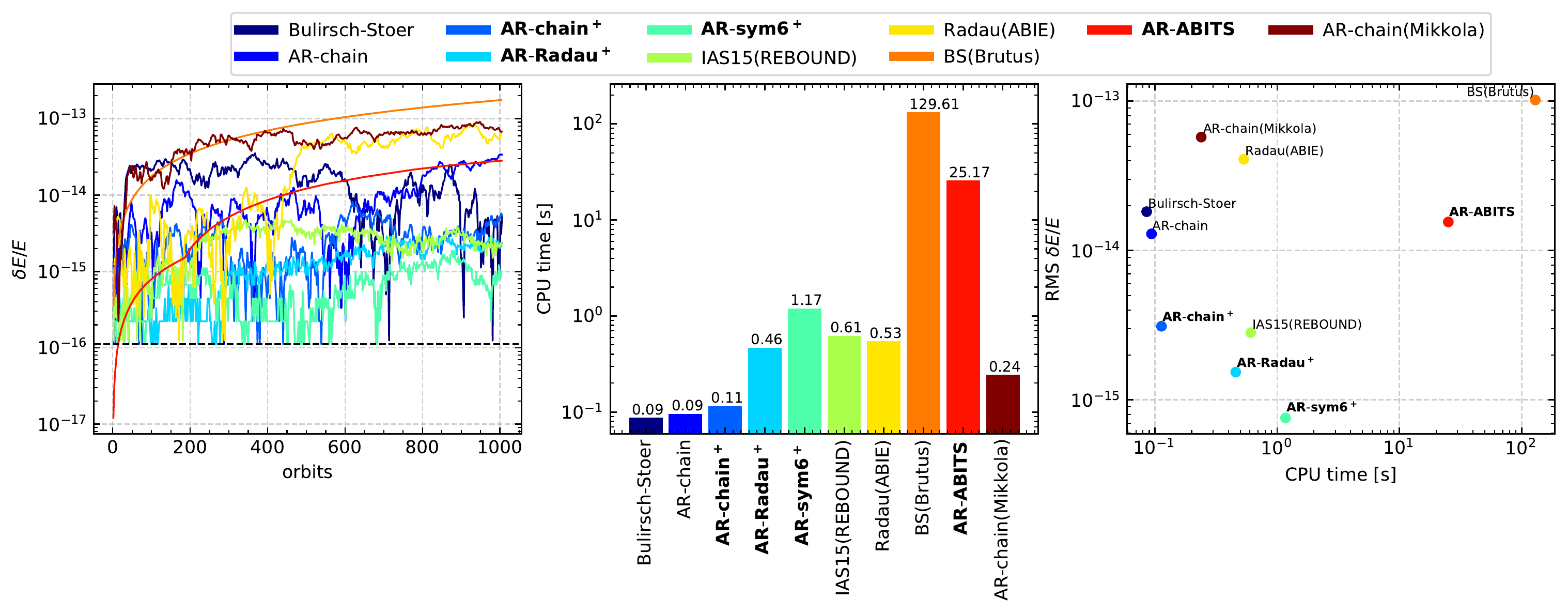}
    \caption{Similar test to Figure~\ref{fig:earth} but applied to the {\bf outer solar system}. The system consists of a central body with a total mass of the inner solar system and four outer planets Jupiter, Saturn, Uranus and Neptune. The integration duration is 1000 Jupiter orbits.}
    \label{fig:outersoalr}
\end{figure*}

From Figure~\ref{fig:outersoalr}, we can see that for this system in which there is no strong interaction, even the BS method in {\tt SpaceHub} can achieve $10^{-14}$. There are not many advantages in introducing the regularization and chain because there is no close encounter or close position subtraction. Therefore, the AR-chain dose not show better precision than the BS method. But both of them show great precision due to the improvements on BS extrapolation. Similarly as in the previous two tests, AR-chain in {\tt SpaceHub} is 2-2.5 times faster than Mikkola's implementation. The AR-chain$^+$ shows one order higher precision than AR-chain in this case because the chain update is not as frequent as in the sun-earth-moon system. Therefore, the active error compensation could do a better job on reducing the round off error.  AR-Radau, AR-sym6 perform slightly better than IAS15 in this case, but AR-sym6 is 2 times slower due to its extra evaluation for error estimation.

The arbitrary precision AR-ABITS and {\tt Brutus} with 88 mantissa bits show significant error accumulation in the test case.  These cases become inefficient in both precision and speed. However, they are designed to achieve extremely high precision by adopting double precision machine precision. One can always achieve higher precision by using more mantissa bits.

\subsubsection{Lidov-Kozai system}\label{sec:test-lk}

In this section, we move on to quantifying the performance of the different integration methods considered in this system for a three-body system undergoing Lidov-Kozai oscillations.

We have already tested extremely eccentric systems, but these are only  two-body systems and some algorithms offer advantages tailored for solving the two-body problem more precisely. To make our tests more robust, we now test a hierarchical three-body system undergoing strong Lidov-Kozai cycles. The system consists of an inner binary with component masses $m_1 = 1 M_\odot$ and $m_2 = 1M_\odot$ and the initial orbital parameters $a_1 = 10$ au and $e_1 = 10^{-3}$. The mass of the outer tertiary is $m_3= 1M_\odot$ with $a_2 = 100$ au and $e_2= 0.5$ initially. The initial inclination angle between the inner and outer orbital planes is $i_{\rm tot}=96.7^\circ$. The maximum $e_1(t)$ reached in this system is larger than $0.999999$. We integrate the system up to $10^5$ years, which corresponds to approximately 6 quadrupole LK cycles. 

\begin{figure*}
      \includegraphics[width=\textwidth]{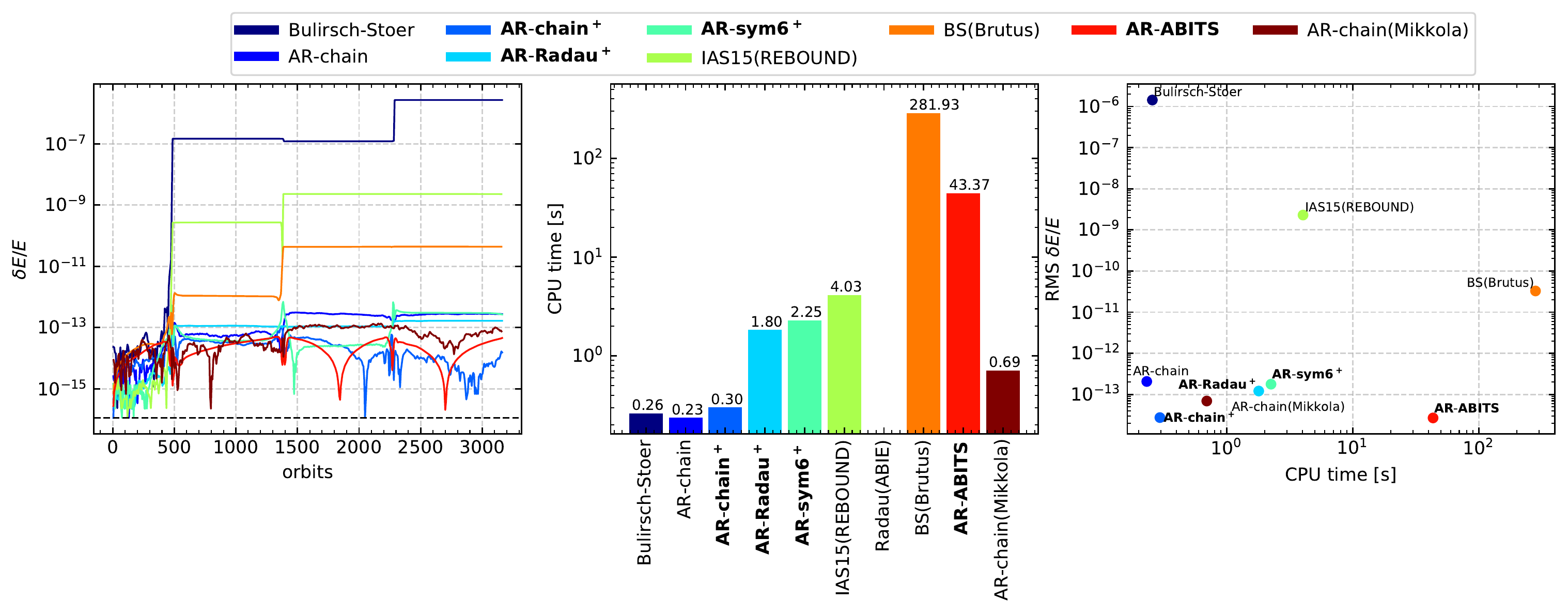}
    \caption{Similar test to Figure~\ref{fig:earth} but for a {\bf Lidov-Kozai system}. The system consists of an inner binary with component masses $m_1 = 1 M_\odot$ and $m_2 = 1M_\odot$ and an initial semi-major axis and eccentricity of, respectively, $a_1 = 10$ au and $e_1 = 10^{-3}$.  The outer tertiary has mass $m_3= 1M_\odot$ with an orbit having an initial semi-major axis and eccentricity of $a_2 = 100$ au and $e_2= 0.5$. The inclination between the inner and outer orbit is 96.7$^\circ$ initially. The integration duration is $10^5$ years, which corresponds to roughly 6 LK cycles.  In this test, the step size of the Radau method in ABIE shrinks to $10^{-14}$ years, implying a very long time to finish the integration test. Thus, ABIE failed in this test.}
    \label{fig:lk}
\end{figure*}

From Figure~\ref{fig:lk}, we see that this three-body system is challenging to model precisely and accurately for some methods. During the integration, the step size of the Radau method in {\tt ABIE} shrinks to $10^{-14}$ years as the eccentricity of the inner binary reaches its maximum value.  Consequently, it takes an excessively long time for the integration to complete. The IAS15 in {\tt Rebound} cannot maintain its high precision after the first eccentricity excitation. The precision drops to $10^{-10}$ for this test case. However, we find that methods  with regularization, including AR-Radau and AR-sym6$^+$, can maintain an error of $\sim 10^{-13}$. The AR-chain method from Mikkola is three times slower. The arbitrary precision integration method characteristic of AR-ABITS in {\tt SpaceHub} behaves much better that than {\tt Brutus} in terms of both speed and precision.

\subsection{Long time integration}

For our final test case, we perform a long term integration test using the new algorithms in {\tt SpaceHub}. We integrate the same system as described in Section~\ref{sec:test-lk} up to 5$\times$10$^9$~years, which corresponds to 1.5$\times 10^8$ orbits of the inner binary and 3 million quadrupole LK cycles. The maximum eccentricity of the system is higher than 0.999999. 

\begin{figure*}
      \includegraphics[width=0.9\textwidth]{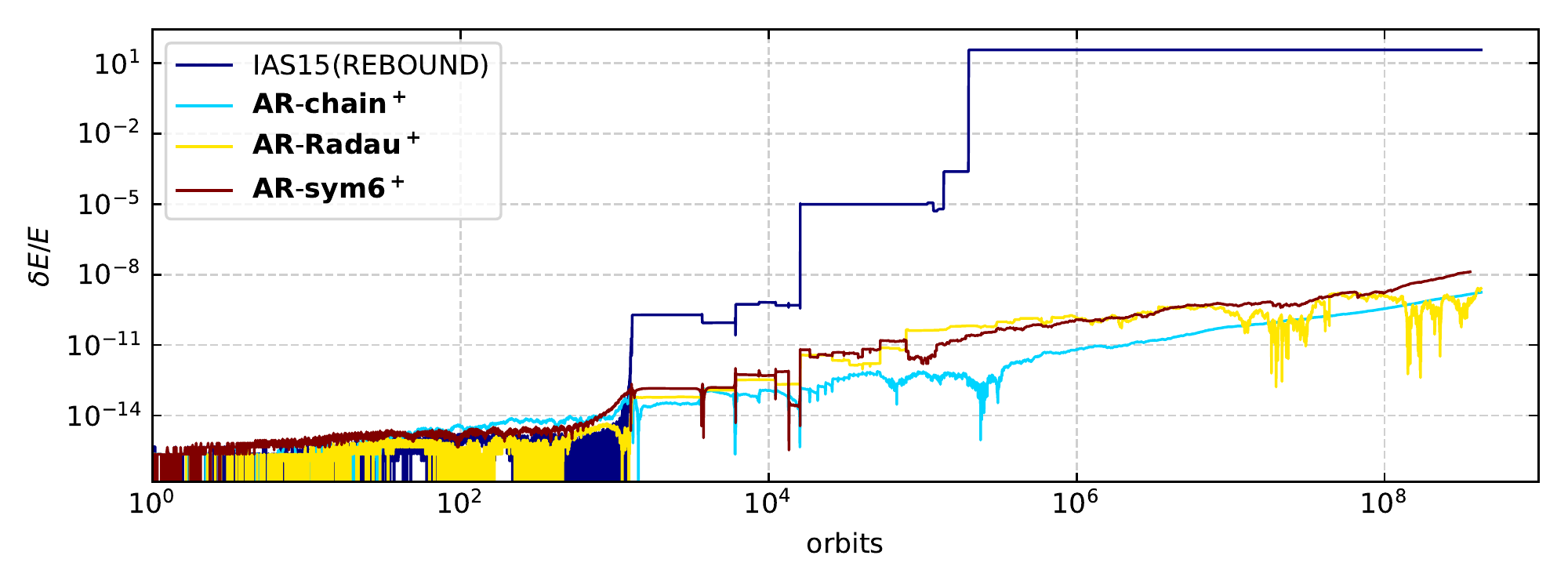}
    \caption{An integration of the same Lidov-Kozai system as in Figure~\ref{fig:lk} but up to $5\times10^9$ years, which is about 1.6$\times10^8$ inner orbits periods. This test is performed on Stony Brook's Seawulf cluster on an Intel Xeon Gold 6148 CPU with GCC-9.2.0 compiler.}
    \label{fig:long-lk}
\end{figure*}

From Figure~\ref{fig:long-lk}, we see that the relative energy error of IAS15 instantly increases around 1000 orbits, where the eccentricity of the inner orbit of the hierarchical triple reaches its first maximum in the first LK cycle. For regularized algorithms that include AR-Chain+, AR-Radau+ and AR-sym6+ in {\tt SpaceHub}, the relative energy error is maintained at 10$^{-13}$. As the integration continues, IAS15 has several significant error jumps due to the high eccentricity of the LK cycles. It becomes completely unreliable around 10$^5$ orbits as the relative energy error grows to 1. The relative energy error of the regularized algorithms in {\tt SpaceHub} accumulate from largely reduced but inevitable round-off errors. For this case, the AR-Radau+ and AR-sym6+ behave similar to AR-Chain+, which was unexpected as we anticipated the latter method to perform better. Indeed, with some extra exploration, we find that in most of the cases, the relative energy errors are at the same order in the long term integration. It depends on the individual nature of specific problems, but the chain algorithm and active error compensation have significantly different impacts on the long term round-off error. Thus, these method can standout in our tests by $\sim$ 1 order of magnitude. However, they all can maintain higher precision and do a better job than other integrator schemes in long term integrations for extremely high eccentricity systems.

\section{Implementation of pair-wise external forces}

In this section, we describe \texttt{SpaceHub}'s implementation of all non-Newtonian forces, including tidal dissipation and Post-Newtonian corrections to account for general relativistic effects.

\subsection{Static tidal forces}

Here we describe our implementation of equilibrium tidal forces based on the weak friction model, where the tides are assumed to take on an equilibrium shape with a constant time lag. In this model, the tidal force exerted on body $m_i$ with radius $R_i$ is implemented as \citep{Hut1981}
\begin{eqnarray}
    \mathbf{F} &=& -3G\frac{m_j^2}{r^2}\bigg( \frac{R_i}{r}\bigg)^5k\bigg(1+3\frac{\dot{r}}{r}\tau\bigg)\hat{\mathbf{r}}
\end{eqnarray}
where $r$, $k$ and $\tau$ are, respectively, the apsidal motion constant and lag time, the relative distance between $m_j$ and $m_i$, the tidal apsidal motion constant and the tidal time lag. 

Then, the acceleration $a_i$ due to the tidal dissipation exerted by body $m_i$ on body $m_j$, and vice versa, can be written
\begin{eqnarray}
    \mathbf{a}_i &=& \mathbf{F}_{\rm tid}/m_j\\
    \mathbf{a}_j &=& \mathbf{F}_{\rm tid}/m_i\,.
\end{eqnarray}
For the tidal force in the radial direction $F_r$, the angular momentum of the two body system is conserved as 
\begin{equation}
    h = \mu \sqrt{GMa(1-e^2)}\,,
\end{equation}
where $\mu = m_im_j/(m_i+m_j)$ is the reduced mass of the two-body system, $M=m_i+m_j$ is the total mass, $a$ is the semi-major axis of the orbit and $e$ is the eccentricity. There are two effects included in this tidal force, the first is the tidal dissipation of the orbital energy, where the orbital energy 
\begin{equation}
    E_{\rm orb} = -G\frac{m_im_j}{2a}
\end{equation}
dissipates at a rate of
\begin{eqnarray}\label{eq:tide-E}
\dot{E}_{\rm orb} = -\frac{9}{2}G^2(m_i+m_j)m_j^2R_i^5k\tau a^{-9} (1-e^2)^{-15/2}e^2f_1(e^2)
\end{eqnarray}
where
\begin{equation}
    f_1(e^2) =1+\frac{15}{2}e^2 +\frac{15}{8}e^4 +\frac{5}{64}e^6\,.
\end{equation}
Due to conservation of angular momentum, one obtains $\dot{a}(t\rightarrow +\infty) = \dot{e}(t\rightarrow +\infty) =0$ and 
\begin{eqnarray}
e(t\rightarrow +\infty) &=&0\\
a(t\rightarrow +\infty) &=&a_0(1-e_0^2)
\end{eqnarray}
where $a_0$ and $e_0$ are the initial semi-major axis and eccentricity, respectively.

The second effect due to this force is the periastron percession, where the pericentre will precess at a rate given by
\begin{equation}\label{eq:tide-omega}
   \dot{\omega}_{\rm tide}= \frac{15}{2}\frac{m_j}{m_i}\frac{R_i^5}{a^5}kn\frac{1+\frac{3}{2}e^2+\frac{1}{8}e^4}{(1-e^2)^5}
\end{equation}
where $n=\sqrt{G(m_i+m_j)/a^3}$ is the mean motion of the binary orbit.
\begin{figure*}
      \includegraphics[width=.49\textwidth]{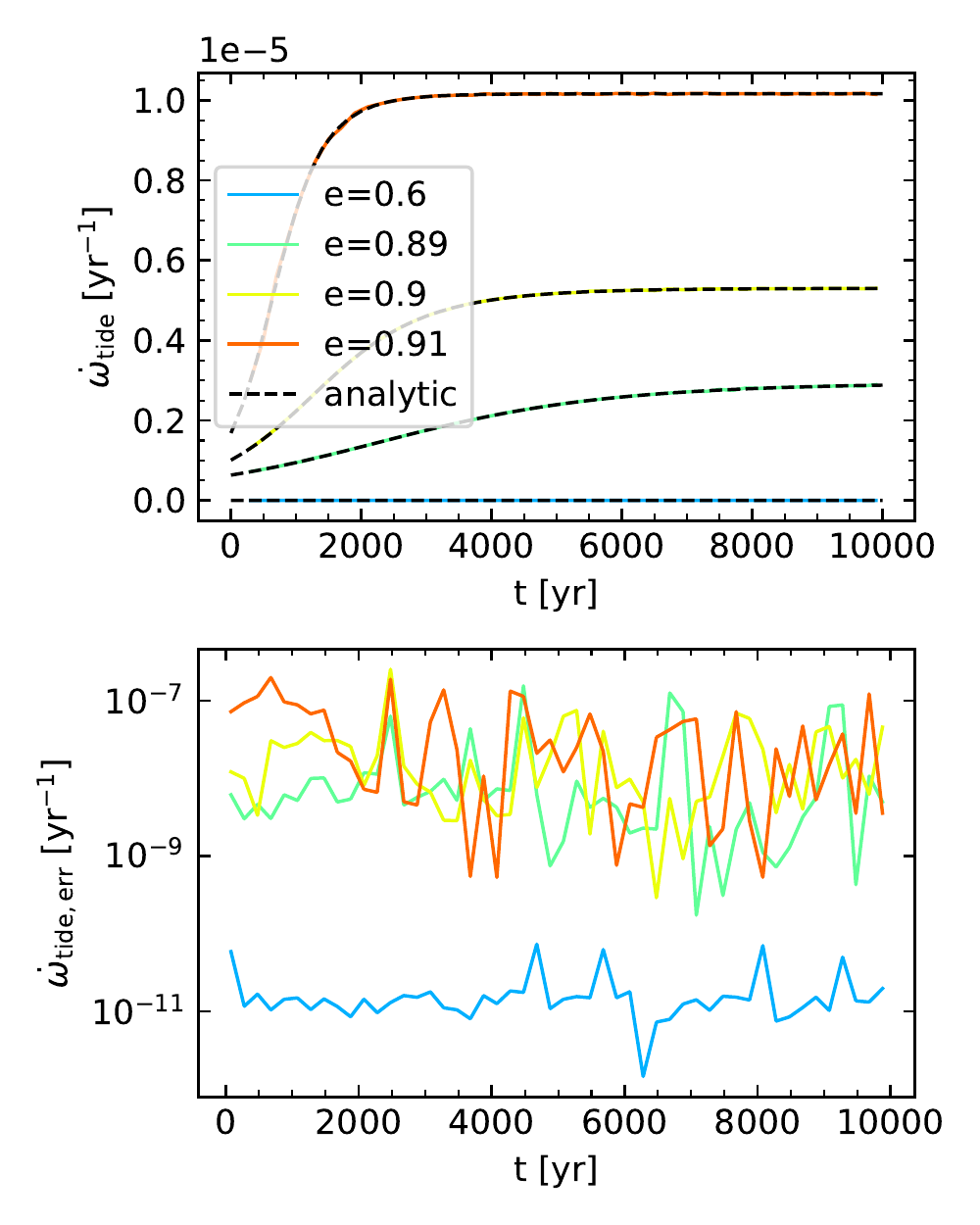}
      \includegraphics[width=.49\textwidth]{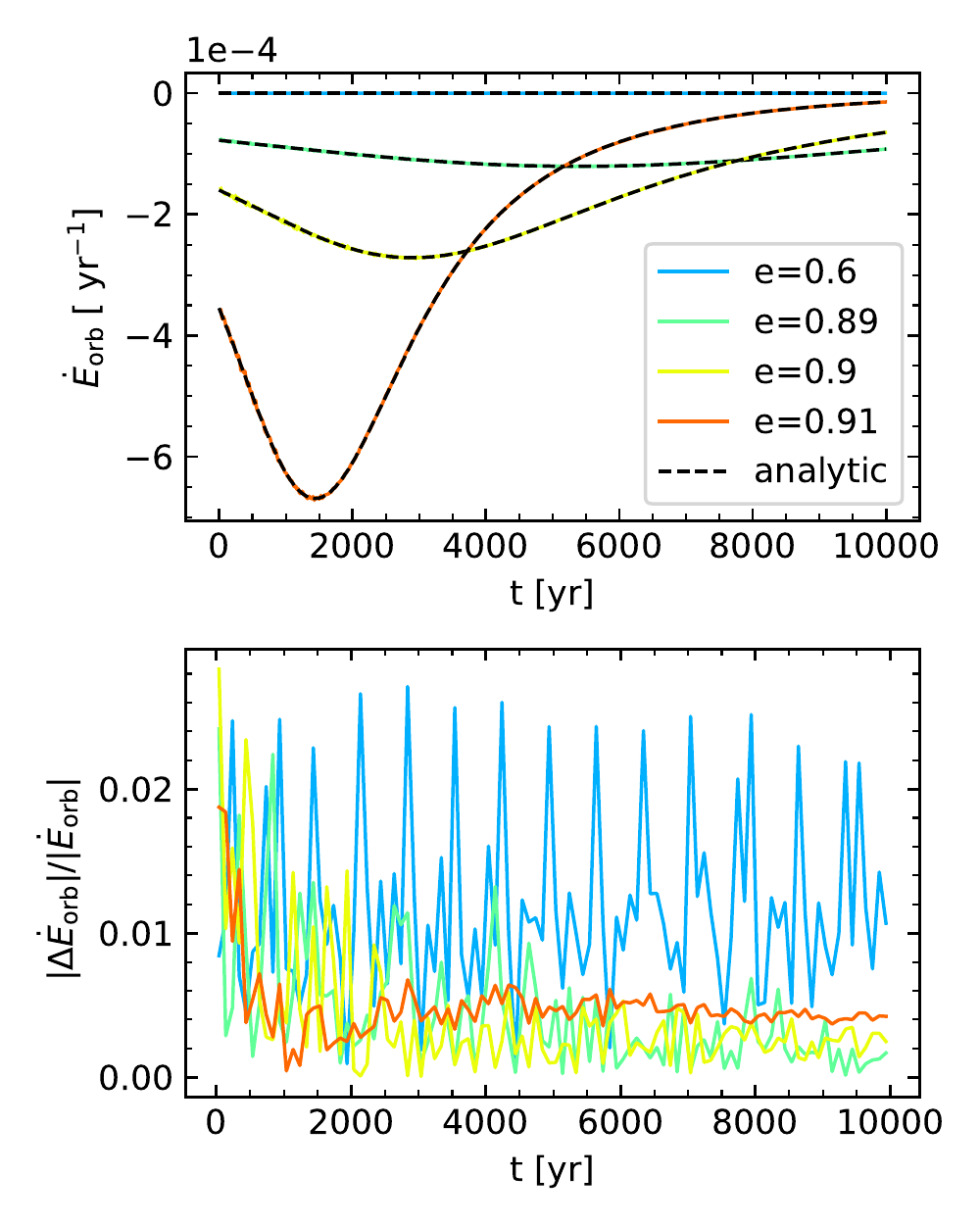}
    \caption{Test of the pericenter precession and orbital energy dissipation from the static tidal force. The system consists of two 1$M_\odot$ stars in a binary with semi-major axis 1 au. Different eccentricities are selected, specifically  0.6, 0.89, 0.9 and 0.91. The primary star is treated as a point mass particle that does not exert tides, while the secondary star exerts tides with apsidal motion constant $k=0.75$ and time lag $\tau=0.25$ years. \textit{Left panels:} Pericenter precession rate for different eccentricities obtained from the simulations and compared to the analytic results obtained from  Equation~\ref{eq:tide-omega}. \textit{Right panels:} Orbital energy dissipation rate calculated from the simulations and compared to the analytic approximation given by Equation~\ref{eq:tide-E}.   }
    \label{fig:tides}
\end{figure*}

Figure~\ref{fig:tides} shows examples of the orbital evolution including the tidal force, where the binary consists of two 1 $M_\odot$ mass stars. The primary star is treated as a point mass, while the secondary star has a radius of 1 $R_\odot$ with $k=0.75$ and $\tau=0.25$~years. The initial semi-major axis of the binary is 1 au, and we test different eccentricities, specifically 0.6 to 0.89, 0.9 and 0.91.

The upper left panel of Figure~\ref{fig:tides} shows the pericentre precession rate for different initial eccentricities as a function of time. We see that the numerical result agrees closely with the analytic prediction given by Equation~\ref{eq:tide-omega}. The bottom left panel shows the absolute difference between the numerical and the analytic results. The upper right panel of Figure~\ref{fig:tides} shows the orbital energy decay rate due to tidal dissipation calculated form the numerical simulations and compared to the analytic results obtained from Equation~\ref{eq:tide-E}. The bottom right panel shows the corresponding relative error of the dissipation rate.

\subsection{Post-Newtonian corrections and general relativistic effects}

In this section, we describe \texttt{SpaceHub}'s treatment of general relativistic effects via the inclusion of Post-Newtonian terms in our estimates for the gravitational acceleration.

The Post-Newtonian approximation in general relativity is of the general form
\begin{equation}
    F_{\rm GR} = c^{-2}F_{\rm 1PN} +c^{-4}F_{\rm 2PN}+c^{-5}F_{\rm 2.5PN}
\end{equation}
where $c^{-2}F_{\rm 1PN}$ contributes most of the periastron precession motion, $c^{-4}F_{\rm 2PN}$ contributes a correction proportional to $(v/c)^2$ and $c^{-5}F_{\rm 2.5PN}$ contributes almost all of the gravitational radiation. The force  exerted on particle $i$ is \citep{Damour1985,Soffel1989},
\begin{eqnarray}
 F_{\rm 1PN} &=& \frac{Gm_im_j}{r^2}\bigg\{  \mathbf{n}\bigg[ -v_i^2 - 2v_j^2 + 4v_iv_j + \frac{3}{2}(nv_j)^2 \nonumber\\
 && + 5\frac{Gm_i}{r} + 4\frac{Gm_j}{r}\bigg] + \mathbf{v} [4nv_i-3nv_j] \bigg\}
 \end{eqnarray}
 \begin{eqnarray}
 F_{\rm 2PN} &=& \frac{Gm_im_j}{r^2}\bigg\{ \mathbf{n}\bigg[-2v_j^4+4v_j^2(v_iv_j) - 2(v_iv_j)^2+\frac{3}{2}v_i^2(nv_j)^2 \nonumber\\
 &&+ \frac{9}{2} v_j^2(nv_j)^2 -6(v_iv_j)(nv_j)^2  -\frac{15}{8}(nv_j)^4 \nonumber\\
 &&+ \frac{Gm_i}{r}\bigg(-\frac{15}{4}v_i^2 +\frac{5}{4}v_j^2-\frac{5}{2}v_iv_j +\frac{39}{2}(nv_i)^2 \nonumber\\
 &&-39(nv_i)(nv_j)+\frac{17}{2}(nv_j)^2\bigg) +\frac{Gm_j}{r}\bigg(4v_j^2 - 8v_iv_j \nonumber\\
 &&+ 2(nv_i)^2 - 4(nv_i)(nv_j) - 6(nv_j)^2\bigg) \bigg] \nonumber\\
 &&+ \mathbf{v}\bigg[ v_i^2(nv_j) + 4v_j^2(nv_i) -5v_j^2(nv_j)-4(v_iv_j)(nv_i) \nonumber\\
 &&+4(v_iv_i)(nv_j) - 6(nv_i)(nv_j)^2 + \frac{9}{2}(nv_j)^3 \nonumber\\
 &&+ \frac{Gm_i}{r}\bigg(-\frac{63}{4}nv_i + \frac{55}{4}nv_j \bigg) +\frac{Gm_j}{r}\bigg(-2nv_i-2nv_j\bigg) \bigg]\nonumber\\
 &&+ \frac{G^2}{r^2}\mathbf{n}\bigg(-\frac{57}{4}m_i^2-9m_j^2-\frac{69}{2}m_im_j\bigg) \bigg\}
 \end{eqnarray}
 
 \begin{eqnarray}
 F_{\rm 2.5PN} &=& m_i\frac{4}{5}\frac{G^2m_im_j}{r^3}\bigg\{\mathbf{n}(nv)\bigg[3v^2 - 6\frac{Gm_i}{r}+\frac{52}{3}\frac{Gm_j}{r}\bigg]\nonumber\\
 &&+\mathbf{v}\Bigg[-v^2 + 2\frac{Gm_i}{r} - 8 \frac{Gm_j}{r}\bigg]\bigg\}
\end{eqnarray}
where $\mathbf{n} = \hat{\mathbf{r}}$ is the unit vector pointing from particle $j$ to particle $i$, and $\mathbf{v} = \mathbf{v}_i - \mathbf{v}_j$ where $\mathbf{v}_i$ and $\mathbf{v}_i$ are the velocities of particles $i$ and $j$, respectively. For simplicity, we have denoted the dot product of the two vectors $\mathbf{x}_1$ and $\mathbf{x}_2$ as $x_1x_2$. To obtain the force exerted on particle $j$, we simply exchange the subscripts $i$ and $j$ in the above equations. Note that the direction of $\mathbf{n}$ and $\mathbf{v}$ will change as well.

\begin{figure}
      \includegraphics[width=.45\textwidth]{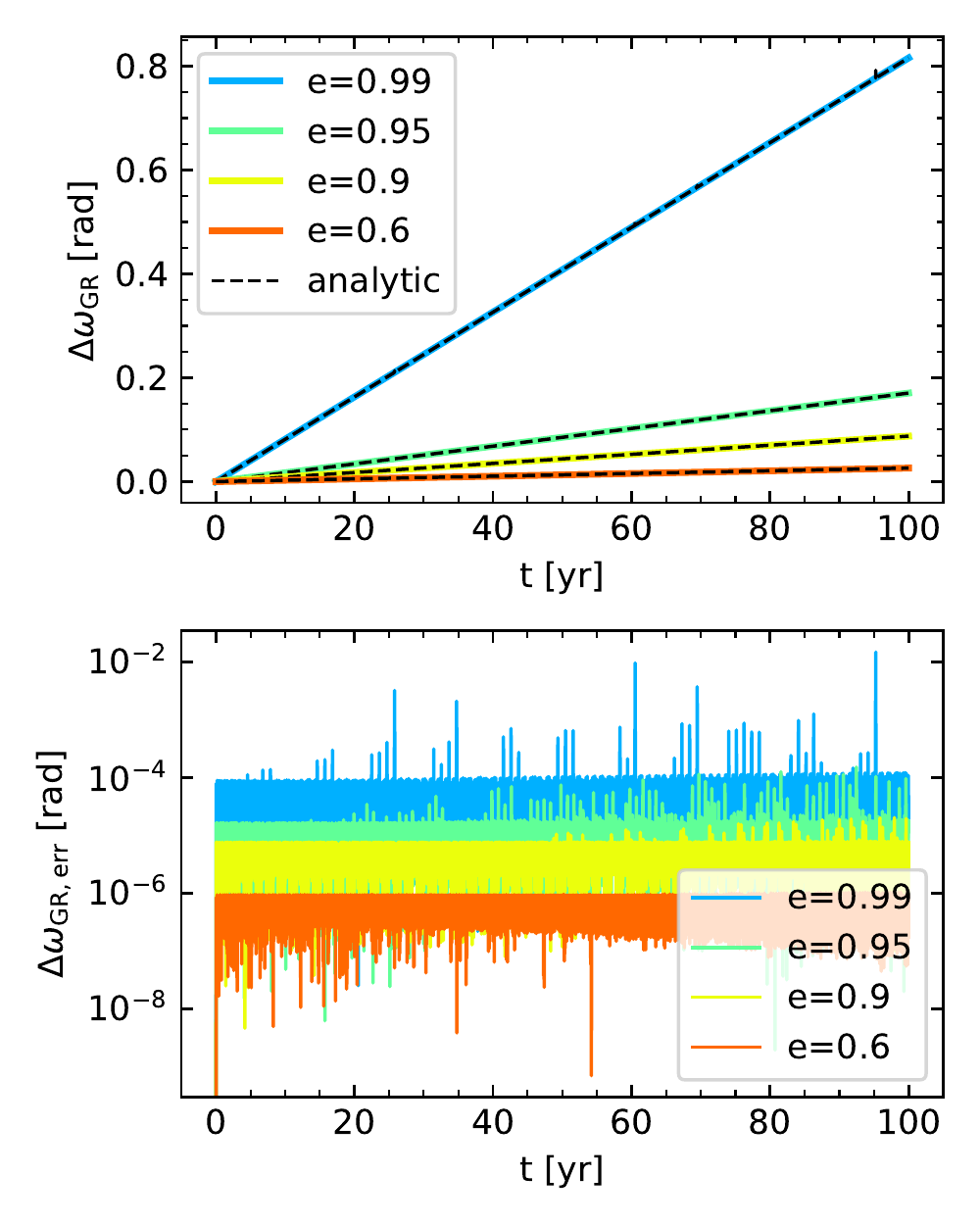}
    \caption{Test of the general relativistic precession adopting only the first order Post-Newtonian term. The binary consists of two identical solar mass stars with an initial semi-major axis of 0.1 AU. The eccentricities are initially set to 0.6, 0.9, 0.95 and 0.99, as indicated by the different colours in the insets. {\it Upper panel:} Calculation results from {\tt SpaceHub} using the AR-Chain$^+$ method.  The dashed black line shows the theoretical expectation for each value of the eccentricity  as described by Equation~\ref{eq:GR}.  {\it Bottom panel:} The relative difference between the results of our computations and the analytic expectation, as a function of time.}
    \label{fig:GR-precession}
\end{figure}

Figure~\ref{fig:GR-precession} shows an example of the time evolution of the relative orbital phase for an identical solar mass binary with an initial semi-major axis of 0.1 AU. We include Post-Newtonian terms up to first order, and observe the resulting pericentre advance due to GR precession.  The solid lines in the upper panel show the precession angle as a function of time assuming different eccentricities. The dashed line is calculated from the analytic equation
\begin{equation}\label{eq:GR}
    \Delta\omega_{\rm GR, analytic} = \frac{24\pi^3a^2}{T^2c^2(1-e^2)}\frac{t}{T}
\end{equation}
where $T$ is the period of the binary and $c$ is the speed of light. The bottom panel shows the difference between the analytic approximation and our simulation results as a function of time. Note that the relative difference between the two remains bounded and does not grow in time. 

\begin{figure*}
      \includegraphics[width=.49\textwidth]{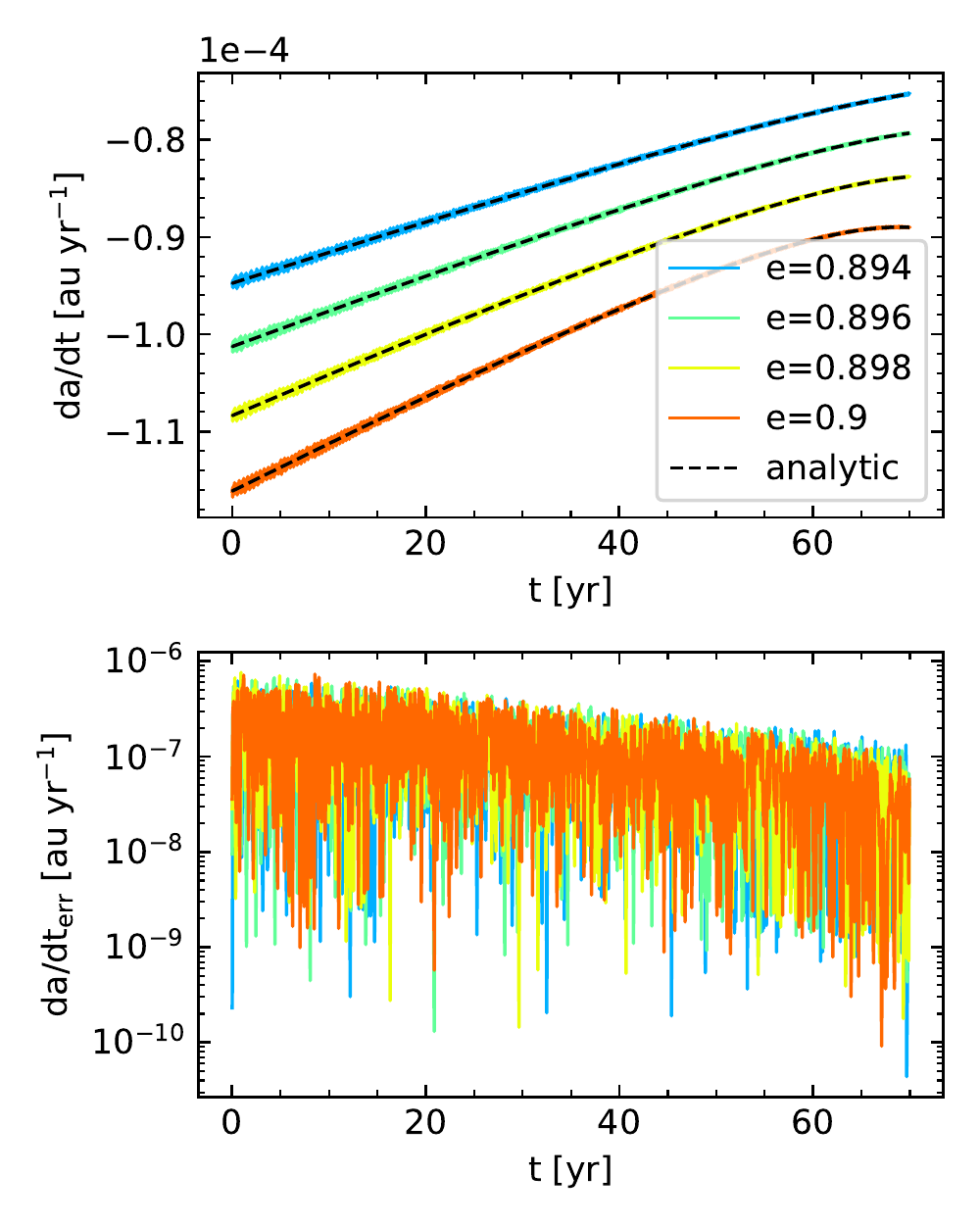}
        \includegraphics[width=.49\textwidth]{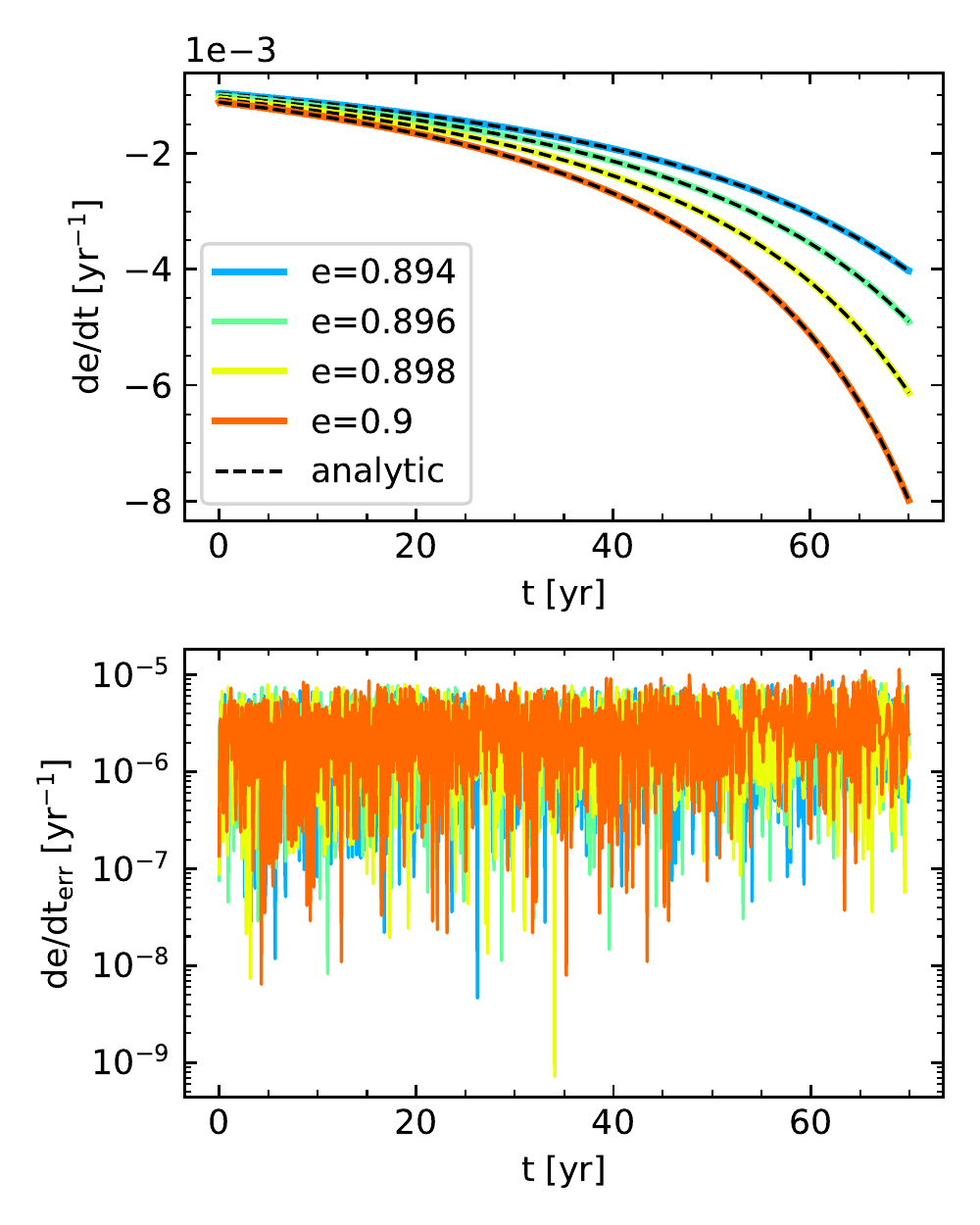}
    \caption{Test of the orbital evolution due to gravitational wave radiation using up to 2.5th order in the Post-Newtonian terms. The binary has component masses of 30 M$_\odot$ and 50 M$_\odot$ with an initial semi-major axis equal to 0.01 AU. The eccentricities are initially set to 0.894, 0.896, 0.898 and 0.9, as indicated by the colour scheme defined in each inset. {\it Upper panels:} Calculation results from {\tt SpaceHub} using the AR-Chain$^+$ method and  the analytic results calculated from Equations~\ref{eq:GW_dadt} and \ref{eq:GW_dedt}.{\it Bottom panels:} The relative difference between the results of our computations and the analytic expectation, as a function of time.}
    \label{fig:GR-radiation}
\end{figure*}

Figure~\ref{fig:GR-radiation} shows the time evolution of the orbital parameters for a tight eccentric black hole binary with component masses $M_1 = 30M_\odot$ and $M_2= 50M_\odot$ and an initial semi-major axis equal to 0.01 AU. We include Post-Newtonian terms up to 2.5th order and observe the subsequent evolution driven by gravitational wave radiation.  The solid lines in the upper panels show the orbital decay rate as a function of time. The bottom panels show the relative difference between our simulated results and the analytic approximation.  We see that the relative difference in the semi-major axis remains less than $10^{-6}$ AU/year $\sim$ 5 mm/s, and the relative difference in the eccentricity remains less than $10^{-5}$/ year. The analytic approximation cosely follows Peter's Equation \citep{peters_gravitational_1964}.

\begin{eqnarray}
    \frac{da}{dt}\bigg|_{\rm GW} &=& -\frac{64}{5}\frac{G^3m_1m_2(m_1+m_2)}{c^5a^3(1-e^2)^{7/2}}(1+\frac{73}{24}e^2+\frac{37}{96}e^4)\label{eq:GW_dadt}\\
    \frac{de}{dt}\bigg|_{\rm GW} &=& -\frac{304}{15}\frac{G^3m_1m_2(m_1+m_2)e}{c^5a^4(1-e^2)^{5/2}}(1+\frac{121}{304}e^2).\label{eq:GW_dedt}
\end{eqnarray}

\section{Conclusions}
We have developed the deeply optimized high precision open source few-body toolkit {\tt SpaceHub}. In this code, several state-of-the-art algorithms are provided that are applicable to a variety of astrophysical few-body problems. The new algorithms include: an algorithmic regularization chain with active round off error compensation AR-Chain+, a regularized arbitrary precision algorithm AR-ABITS, a regularized higher order symplectic method with active round off error compensation AR-sym6+ and a regularized Gauss-Radau method with active round off error compensation AR-Radau+. By comparing to popular high precision few-body codes via various applications to the time evolution of various astrophysical test cases, we show that {\tt SpaceHub} consistently provides the most precise, accurate and fastest algorithm for most specific astrophysical problems of interest in the few-body limit. 

We begin by briefly reviewing the existing Bulirsch-Stoer-based high precision integration methods and Gauss-Radau-based methods, and discuss the improvements we have made on these algorithms.  In Section~\ref{sec:arabits}, we discuss the arbitrary precision method with extended floating point precision. We then go on to discuss our implementation and improvements in our new regularized arbitrary precision method AR-ABITS. The AR-ABITS method achieves arbitrary precision based on the GBS extrapolation. By adopting an optimal extrapolation step sequence and a fine-tuned extrapolation process, the round off error can be significantly reduced with the provided bits floating-point numbers. We show that to achieve the same arbitrary precision, AR-BITS is roughly 1-2 orders of magnitude faster than the popular arbitrary precision code {\tt Brutus}. Apart from this, we introduce regularization into the arbitrary precision method, which makes it even more efficient in dealing with highly eccentric systems.

In section~\ref{sec:archain+}, we discuss the original algorithmic regularization chain algorithm AR-chain. In this section, we propose an improved chain coordinate transformation that eliminates the centre-of-mass reduction.  This saves non-negligible CPU time in the few-body regime, and introduces the active round off error compensation into the AR-chain to form the AR-chain+ method. The new algorithm is faster and more accurate in the high precision regime, where the round off error is non-negligible.

In Section~\ref{sec:arsym6+}, we discuss regularization in higher order symplectic methods. The AR-sym6+ algorithm, a sixth-order regularized symplectic method with active round off error compensation, makes it possible to accurately and efficiently solve extremely eccentric systems and very close pair-wise encounters with fixed step size.  Consequently, the symplectic nature of the evolving system is preserved.

In Section~\ref{sec:arradau+}, we introduce regularization into the Gauss-Radau method using extended general coordinates. With the regularization and active round off error compensation, the AR-Radau+ algorithm becomes more efficient in solving the time evolution of extremely eccentric orbits than the original Gauss-Radau method. At the same time, it preserves the advantages of the original method in long time integrations. 

{\tt SpaceHub} is fully open source. All of the new state-of-the-art algorithms discussed above can be accessed via GitHub at \url{https://yihanwangastro.github.io/SpaceHubWeb/}. Together with the implementation of additional pair-wise interactions, such as tidal and Post-Newtonian forces, not to mention a myriad of other performance and optimization tools, {\tt SpaceHub} undoubtedly competes with, challenges and even surpasses the most commonly used codes and gravity integrators used in the field today for dealing with astrophysical problems ranging from extrasolar planetary systems, black hole binaries, etc., in terms of not only accuracy and precisions but also speed.

\section*{Acknowledgements}
N.W.C.L. gratefully acknowledges support from the Chilean government via Fondecyt Iniciacion Grant 11180005, and acknowledges financial support from Millenium Nucleus NCN19\_058 (TITANs). 
Bin Liu gratefully acknowledges support from the European Union’s Horizon 2020 research and innovation program under the Marie Sklodowska-Curie grant agreement No. 847523 ‘INTERACTIONS’. RP gratefully acknowledges support from NSF award AST-2006839.

\section*{Data Availability Statements}
Data are available in a repository and can be accessed via \url{https://yihanwangastro.github.io/SpaceHubWeb/}.



\bibliographystyle{mnras}
\bibliography{example} 





\bsp	
\label{lastpage}
\end{document}